\documentclass{emulateapj}
\usepackage{graphicx}

\slugcomment {submitted to the Astrophysical Journal }

\shorttitle{2D Distribution of Iron in SN 1885 (S And)}

\shortauthors{Fesen, H{\"o}flich, and Hamilton}

\newcommand{\unit}[1]{\, {\rm #1}}

\begin{document}

\title{The 2D Distribution of Iron Rich Ejecta in the Remnant of SN~1885 in M31\altaffilmark{1}}

\author{Robert A.\ Fesen\altaffilmark{2},
        Peter A. H{\"o}flich\altaffilmark{3}, and
        Andrew J. S. Hamilton\altaffilmark{4}
        }

\altaffiltext{1}{Based on observations with the NASA/ESA Hubble Space Telescope,
obtained at the Space Telescope Science Institute,
which is operated by the Association of Universities for Research in
Astronomy, Inc.\  under NASA contract No.\ NAS5-26555.}
\altaffiltext{2}{6127 Wilder Lab, Department of Physics \& Astronomy,
                 Dartmouth College, Hanover, NH 03755}
\altaffiltext{3}{Department of Physics, Florida State University, Tallahassee, FL 32306}
\altaffiltext{4}{JILA and the Department of Astrophysical and Planetary Sciences,
                 University of Colorado, Boulder, CO 80309}

\begin{abstract}

We present Hubble Space Telescope ({\sl HST}) ultraviolet \ion{Fe}{1} and
\ion{Fe}{2} images of the remnant of Supernova~1885 (S~And) which is observed
in absorption against the bulge of the Andromeda galaxy, M31.  We compare these
\ion{Fe}{1} and \ion{Fe}{2} absorption line images to previous {\sl HST}
absorption images of S And, of which the highest quality and theoretically
cleanest is \ion{Ca}{2}~H~\&~K.  Because the remnant is still in free
expansion, these images provide a 2D look at the distribution of iron
synthesized in this probable Type~Ia explosion, thus providing insights and
constraints for theoretical SN~Ia models. The \ion{Fe}{1} images show extended
absorption offset to the east from the remnant's center as defined by
\ion{Ca}{2} images and is likely an ionization effect due to self-shielding.
More significant is the remnant's apparent \ion{Fe}{2} distribution which
consists of four streams or plumes of Fe-rich material seen in absorption that
extend from remnant center out to about $10{,}000 \unit{km} \unit{s}^{-1}$.
This is in contrast to the remnant's \ion{Ca}{2} absorption, which is
concentrated in a clumpy, broken shell spanning velocities of $1000 - 5000 \unit{km}
\unit{s}^{-1}$ but which extends out to $12{,}500 \unit{km} \unit{s}^{-1}$.
The observed distributions of Ca and Fe rich ejecta in the SN~1885 remnant are
consistent with delayed detonation white dwarf models.  The largely spherical
symmetry of the Ca-rich layer argues against a highly anisotropic explosion as
might result from a violent merger of two white dwarfs.

\end{abstract}


\section{Introduction}

The prevailing picture of common Type Ia supernovae (SNe~Ia) is that they are
explosions of degenerate carbon-oxygen white dwarfs that undergo a
thermonuclear runaway when they reach the Chandrasekhar limit as a result of
mass transfer in some type of close binary stellar system
\citep{hf60,CK69,Nomoto84,HN00,li03}.  Two categories of progenitor system are
considered promising: a ``single degenerate'' system comprising a single white
dwarf that accretes from a companion main sequence, helium, or red giant star
\citep{branch95,nomoto03,wang2012,Stephano11} and a ``double degenerate''
system in which two white dwarfs merge after losing angular momentum by
gravitational radiation \citep{Iben84,webbink84}.  The diversity of observed
SNe~Ia suggests that more than a single scenario may operate
\citep{hk96,Quimby06,Hillebrandt13}. Possible SN~Ia progenitor systems
have been discussed recently by \citet{Howell2011}, \citet{Nugent2011},
\citet{Bloom2012}, \citet{Stefano2012}, and \citet{Hoeflich2013}.

The mechanism by which the explosion proceeds in white dwarfs remains poorly
understood.  Thermonuclear runaway models involving a pure detonation (a
supersonic shock wave) appear to be ruled out because they predict that the
white dwarf would be nearly entirely incinerated to iron-group elements, chiefly
$^{56}$Ni, whereas observed spectra of SNe~Ia show substantial quantities of
intermediate mass elements.  Similarly, models involving a pure deflagration (a
subsonic burning wave) are ruled out because the convectively unstable
deflagration front effectively mixes all elements radially, whereas observed
spectra and light curves suggest a layered structure with intermediate mass
elements on the outside and nickel-iron on the inside.

\cite{khokhlov91} proposed that SN~Ia light curves could be explained
empirically by a ``delayed detonation'' scenario in which the explosion starts
in the core as a deflagration wave, which is followed somehow by a detonation
wave of overlying unburned material which unbinds the white dwarf completely.
Heat conduction from the initial deflagration front pre-heats and expands the
star's outer layers, lifting their electron degeneracy so that when those outer
layers subsequently detonate, their explosion is softened. The consequence is
that burning of the outer layers does not continue to completion, and a layered
structure of intermediate mass elements is produced like that observed. 

Numerical calculations indicate that the thermonuclear runaway of a near
Chandrasekhar-mass white dwarf is preceded by a century long, highly
turbulent, convective, carbon-burning, ``smoldering'' phase
\citep{Sugimoto80,Nomoto82,HN00}.  Ignition, which occurs at around $1.5 \times
10^{9} \unit{K}$, is highly sensitive to temperature and density and may occur
off-center and proceed along paths of least resistance determined by the
fluctuating conditions around the ignition point
\citep{Timmes1992,Garcia1995,Niemeyer1996,hs02,Livne05}.

Recent three-dimensional numerical simulations suggest that deflagration may
proceed along a single dominant plume that starts off-center and breaks through
to the surface of the white dwarf \citep{Plewa07}.  While hydrodynamical
calculations during the deflagration phase depend sensitively on the initial
conditions at the time of the thermonuclear runaway, some calculations
suggest strong mixing of partially and complete burning products throughout the
WD \citep{gamezo03,roepke06}.
 
SN~Ia explosion computations are quite challenging, involving a complex interplay of
turbulent hydrodynamics, nuclear burning, conduction, radiative transfer in
iron-group rich material, and perhaps magnetic fields leading to 
uncertainties
\citep{khokhlov95,Neimeyer95,Livne99,Reinecke99,gamezo2004,roepke12}.  
Several key questions about expansion asymmetries and the overall characteristics of
SNe~Ia could be resolved if one could obtain direct observations of the distribution and
kinematics of elements in young SN~Ia remnants. 


\begin{deluxetable*}{cccclccl}
\tablecolumns{8}
\tablecaption{Log of Hubble Space Telescope Images of the Remnant of SN~1885 in M31}
\tablehead{
\colhead{{\sl HST}} & \colhead{UT}   & \colhead{Observation} & \colhead{Exposure}
           & \colhead{Filter} & \colhead{Central} &  \colhead{Program} & \colhead{Comment}  \\
\colhead{Instrument} & \colhead{Date} & \colhead{ID} &
\colhead{Time (s)}   & \colhead{ID} & \colhead{Wavelength} & \colhead{ID} & \colhead{}}
\startdata
ACS/HRC & 2004-11-20   & J8ZS05010 & 15600  & F250W  & $2716 \unit{\AA}$ & 10118  & \ion{Fe}{1}, \ion{Fe}{2}, \ion{Mg}{1} \& \ion{Mg}{2} \\
ACS/HRC & 2004-10-13   & J8ZS04010 & 5200   & F330W  & $3363 \unit{\AA}$ & 10118  & UV continuum   \\
ACS/WFC & 2009-12-14   & JB2701010 & 12700  & FR388N & $3724 \unit{\AA}$ & 11722  & \ion{Fe}{1} $3720 \unit{\AA}$ \\
ACS/WFC & 2004-11-08   & J8ZS03010 & 8960   & FR388N & $3951 \unit{\AA}$ & 10118  & \ion{Ca}{2} H \& K  \\
ACS/WFC & 2004-11-02   & J8ZS02010 & 8960   & FR423N & $4227 \unit{\AA}$ & 10118  & \ion{Ca}{1} $4227 \unit{\AA}$ \\
ACS/WFC & 2004-08-25   & J8ZS01010 & 4668   & FR462N & $4600 \unit{\AA}$ & 10118  & blue continuum  \\
WFC3/UVIS    & 2012-11-18   &\ion{}{1}BR801010 & 8064   & F225W   & $2366 \unit{\AA}$ & 12609 & \ion{Fe}{1} \& \ion{Fe}{2} \\
WFC3/UVIS    & 2012-11-21   &\ion{}{1}BR802010 & 8064   & F225W   & $2366 \unit{\AA}$ & 12609 &  ~ "   ~ ~~~~ " \\
WFC3/UVIS    & 2012-12-01   &\ion{}{1}BR803010 & 8064   & F225W   & $2366 \unit{\AA}$ & 12609 &  ~ "   ~ ~~~~ " \\
WFC3/UVIS    & 2010-07-21   &\ion{}{1}BF310040 &  925   & F275W   & $2707 \unit{\AA}$ & 12058 & \ion{Fe}{1}, \ion{Fe}{2}, \ion{Mg}{1} \& \ion{Mg}{2}  \\
WFC3/UVIS    & 2010-07-23   &\ion{}{1}BF311040 &  925   & F275W   & $2707 \unit{\AA}$ & 12058 & ~ "   ~ ~~~~ "  ~ "   ~ ~~~~ "  \\
WFC3/UVIS    & 2010-07-21   &\ion{}{1}BF310030 & 1250   & F336W   & $3355 \unit{\AA}$ & 12058 & UV continuum \\
WFC3/UVIS    & 2010-07-23   &\ion{}{1}BF311030 & 1250   & F336W   & $3355 \unit{\AA}$ & 12058 & ~ "   ~ ~~~~ "  \\
WFC3/UVIS    & 2012-12-01   &\ion{}{1}BR803020 & 2724   & F336W   & $3355 \unit{\AA}$ & 12609 &  ~ "   ~ ~~~~ "  \\
WFC3/UVIS    & 2010-12-21   &\ion{}{1}BIR01020 & 2700   & F373N   & $3730 \unit{\AA}$ & 12174 & \ion{Fe}{1} $3720 \unit{\AA}$   \\
WFC3/UVIS    & 2010-12-25   &\ion{}{1}BIR06020 & 2700   & F390M   & $3897 \unit{\AA}$ & 12174 & \ion{Ca}{2} H \& K \\
WFC3/UVIS    & 2010-12-23   &\ion{}{1}BIR05020 & 2700   & F502N   & $5010 \unit{\AA}$ & 12174 & [\ion{O}{3}] $5007 \unit{\AA}$  \\
WFC3/UVIS    & 2010-12-26   &\ion{}{1}BIR07020 & 2700   & F547M   & $5447 \unit{\AA}$ & 12174 & V band   \\
WFC3/UVIS    & 2010-12-21   &\ion{}{1}BIR02020 & 2700   & F656N   & $6561 \unit{\AA}$ & 12174 & H$\alpha$ $6563 \unit{\AA}$  \\
WFC3/UVIS    & 2010-12-21   &\ion{}{1}BIR03020 & 2700   & F658N   & $6584 \unit{\AA}$ & 12174 & [\ion{N}{2}] $6584 \unit{\AA}$ \\
WFC3/UVIS    & 2011-01-02   &\ion{}{1}BIR08020 & 2700   & F665N   & $6650 \unit{\AA}$ & 12174 & H$\alpha$ off-band \\
\enddata
\label{tab:table}
\end{deluxetable*}


\begin{figure*}
\begin{minipage}{175mm}
\begin{center}
\leavevmode
\includegraphics[scale=.7]{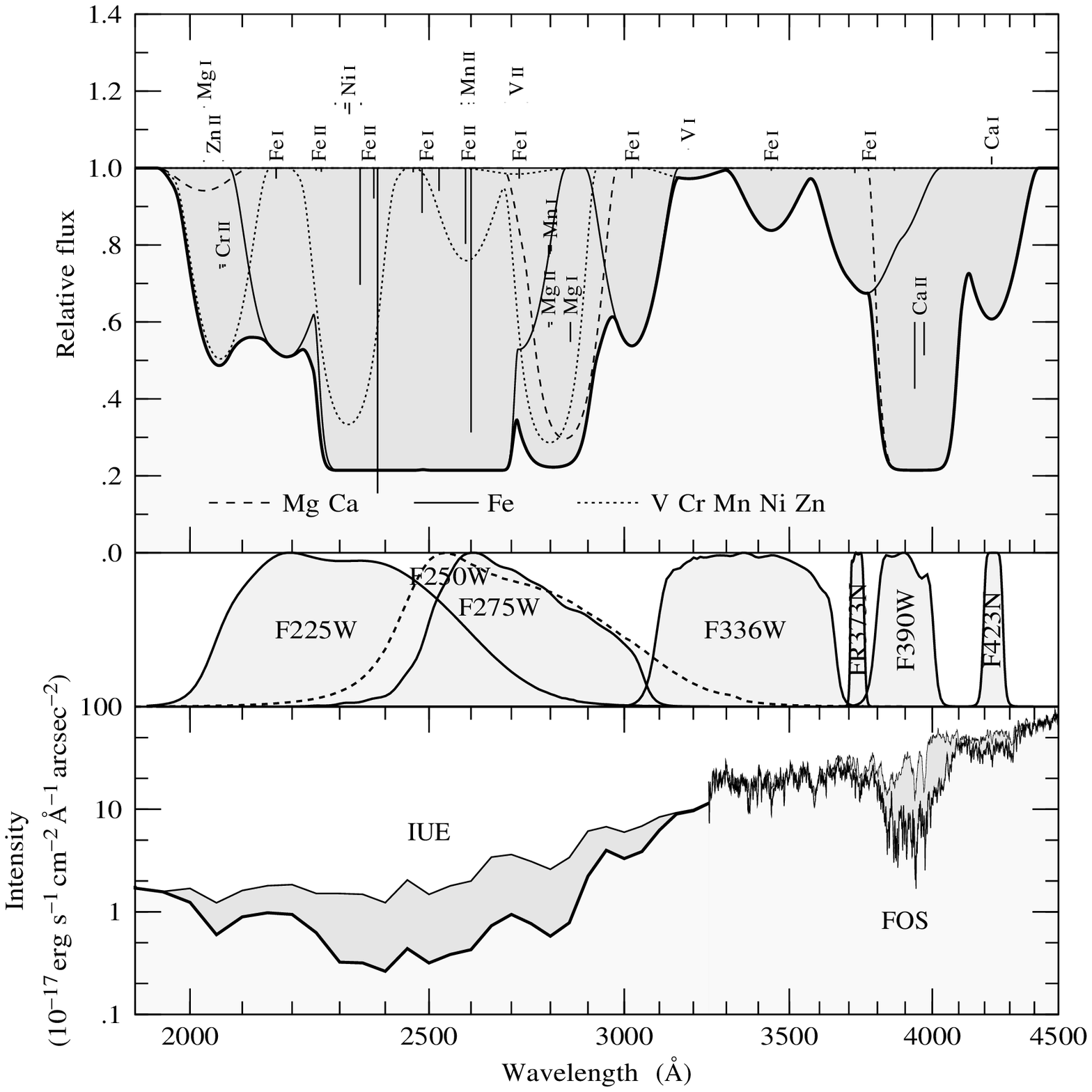}
\end{center}
\caption{
\label{specuv}
{\sl Top:}
Model UV-optical absorption spectrum of SN~1885, from \citet{Fesen1999}
Vertical lines, with lengths proportional to oscillator strengths,
mark wavelengths of the strongest lines.
{\sl Middle:} Bandpasses for the WFC3 F225W,  F275W, F336W, and  F390M filters,
along with the ACS/HRC F250W and ramp FR388N filters, 
are shown with filter peak transmissions normalized to unity.
See text for details.
{\sl Bottom:}
Archival IUE spectrum of the M31 bulge near the location of SNR~1885
\citep{Burstein1988},
along with the {\sl HST}/FOS spectrum of SNR~1885 \citep{Fesen1999}.
The upper line is the bulge spectrum,
while the lower line is the spectrum absorbed through SNR~1885,
expected in the case of IUE, observed in the case of FOS.
}
\end{minipage}
\end{figure*}

However, freely expanding ejecta cool adiabatically, thereby becoming faint and
effectively invisible via line emission in just a few years. Nonetheless, as
long as a remnant's ejecta are still in the free expansion phase the velocity
distribution of elements remains essentially the same as that established
shortly after the explosion.  

Fortunately, such an investigation is possible for SN~1885, the bright
historical nova known as S~Andromeda (S~And).  S And was discovered in late
August of 1885 located at a projected distance just $16''$ away from M31's
nucleus.  Because of its central location in M31, the remnant's expanding
ejecta is visible not by emission, but via resonance line absorption against
the background of the Andromeda galaxy's bulge stars.  SN~1885's reported
optical spectrum lacked hydrogen lines, defining it as Type~I \citep{deV85}.
Its reddish appearance and fast light curve suggests a subluminous Type~Ia
event \citep{deV85} although this classification is uncertain
(see \citealt{CP88,Pastorello08,Perets2011}).

Numerous searches for the remnant of SN~1885 in emission failed for nearly a
century (e.g.\ those of W.\ Baade discussed by \citealt{Osterbrock2001}).
The remnant was finally detected through a ground-based, near-UV image
which revealed a small dark spot of Ca and Fe resonance line absorption
silhouetted against the starry background of M31's bulge \citep{Fesen89}.

Follow-up images taken with the Hubble Space Telescope {\sl HST} revealed a
circular $\simeq 0\farcs75$  diameter dark spot produced by a blend of
\ion{Ca}{2} H \& K  line absorption \citep{Fesen1999}. Subsequent ultraviolet
imaging of the remnant with {\sl HST} using the WFC2 F255W filter revealed a
$0\farcs5$ diameter absorption spot likely due largely to saturated UV
\ion{Fe}{2} resonance lines \citep{Hamilton00}.  

Spectra taken with Faint Object Spectrograph (FOS) on {\sl HST}) established
that the absorption was produced principally by \ion{Ca}{2}~K~\&~H $3934,3968
\unit{\AA}$, with additional contributions from \ion{Ca}{1} $4227 \unit{\AA}$
and \ion{Fe}{1} $3720,3441 \unit{\AA}$ \citep{Fesen1999}.  The remnant's
\ion{Ca}{2} absorption was found to extend to a maximum velocity of $\simeq
13{,}100 \pm 1500 \unit{km} \unit{s}^{-1}$.  The strength of the \ion{Fe}{1}
3720 absorption line, the relative strengths of \ion{Ca}{1} and \ion{Ca}{2}
lines, and the depth of the imaged \ion{Fe}{2} absorption spot were used to
estimate an iron mass (in the form of \ion{Fe}{2}) at between 0.1--1.0
M$_{\odot}$.

Narrow passband images taken in 2004 with the Wide Field Channel of the
Advanced Camera for Surveys (ACS/WFC) on {\sl HST} showed that the \ion{Ca}{2}
absorption is roughly spherical with a maximum diameter of $0\farcs 8$
\citep{Fesen2007}.  At the known distance $785 \pm 25 \unit{kpc}$ of M31
\citep{McConn2005}, this angular diameter corresponds to a mean expansion
velocity of $12{,}500 \unit{km} \unit{s}^{-1}$ over the $\simeq$ 120 yr age of
SN~1885.  The agreement between the remnant's size and its expansion
velocity as measured in the \ion{Ca}{2} absorption implies that the SN~1885
remnant is still virtually in free expansion.  This is also consistent with the
fact that the remnant is an exceptionally weak radio source \citep{SD01,Hof13}
with no confirmed X-ray emission \citep{Kaaret02}.

Since its ejecta are in free expansion, the distribution of elements is
essentially the same as that shortly after the explosion.  The dominant element
near the center of the remnant at the present time is expected to be iron, and
the dominant iron ionization species is expected to be \ion{Fe}{2}.  The
presence of \ion{Fe}{1} in the FOS spectrum attests to the relatively low
ionization state of the supernova ejecta in general. The photoionization
timescale of \ion{Fe}{1} exposed to the observed flux of ultraviolet light from
the bulge of M31 is of order 10~years \citep{Fesen1999}. The fact that
\ion{Fe}{1} is still present can be attributed to partial self-shielding 
by its own continuum optical depth.

Although \ion{Fe}{2} is expected to be the dominant ion of iron, imaging the
remnant's iron-rich material via \ion{Fe}{2} absorption is challenging. This is
because the bulge of M31 is faint at the wavelengths $\sim 2500 \unit{\AA}$
of the strongest near-UV \ion{Fe}{2} absorption lines.  A detection of SN~1885 in
\ion{Fe}{2} using the Wide Field Planetary Camera 2 (WFPC2) on {\sl HST} was
reported by \cite{Hamilton00} but the remnant was barely resolved with a low
signal-to-noise. 

The Wide Field Planetary Camera 3 for UV and optical imaging (WFP3/UVIS)
installed on {\sl HST} in 2009 is significantly more sensitive than its
precursors, WFPC2 and ACS.  Here we present a spatially resolved, long exposure
image of the 2D distribution of \ion{Fe}{2} in SN~1885 obtained with WFPC3.
The WFC3/UVIS filter F225W, covers the strong 2344 and 2383 $\unit{\AA}$ resonance
lines of \ion{Fe}{2} and is largely free of contamination of absorption lines
from lighter elements.  We compare this image to other {\sl HST} images of
SN~1885, including \ion{Fe}{1} images, and we discuss the implications of these
data for explosion models of Type~Ia.  The observations and results are
described in \S2 and \S3, and the implications are discussed in \S4.  A summary
of the findings and conclusions is given in \S5.


\begin{figure*}[t]
   \begin{center}
        \includegraphics[scale=.66]{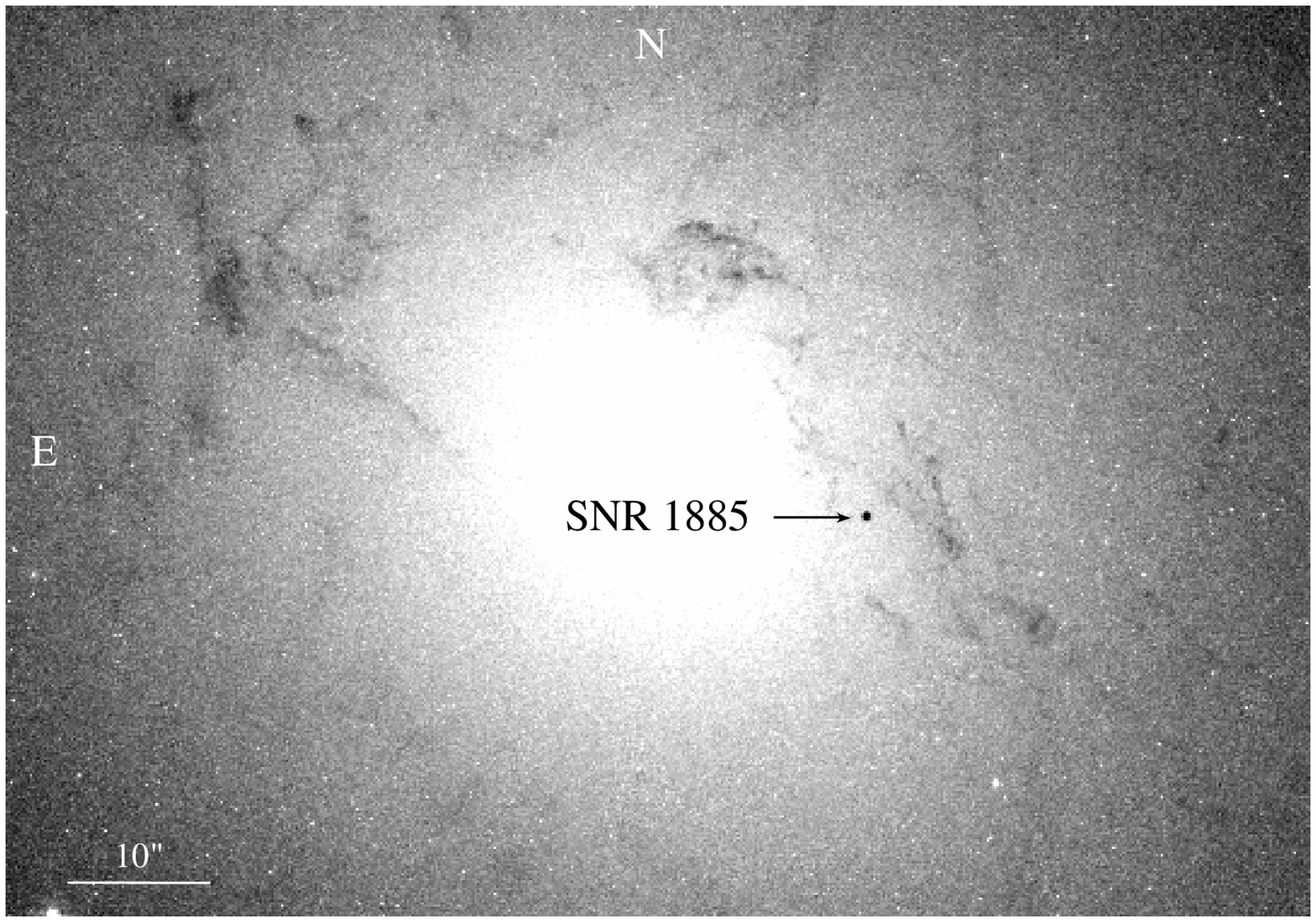}
        \hfill
\includegraphics[scale=.66]{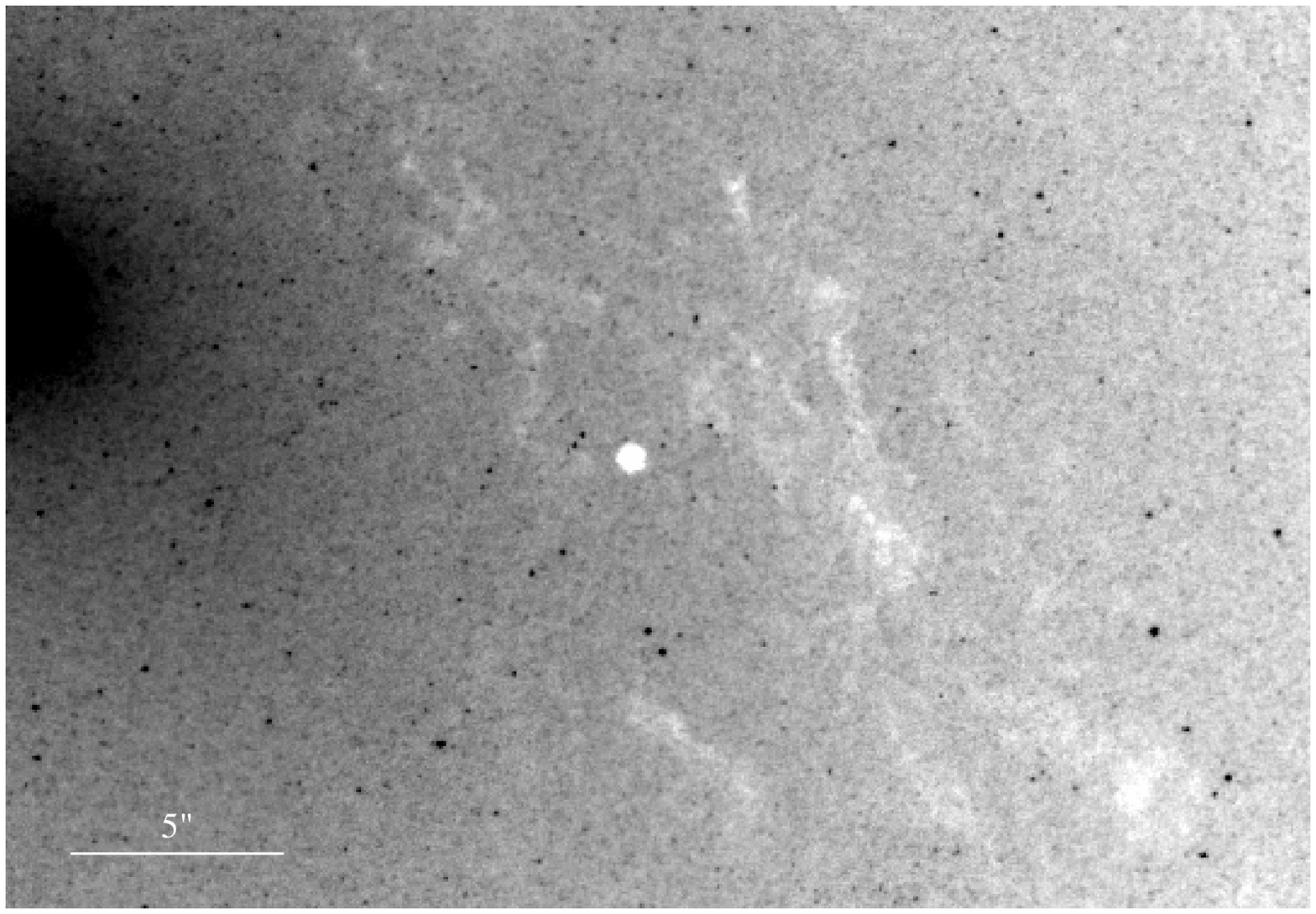}
    \end{center}
        \caption{A December 2010 WFC3 image of the bulge of M31 taken with the F390M filter
as part of an emission line mapping program of the nuclear regions of M31 (PI: Z. Li).
A positive linear stretch of the M31 bulge is shown in the upper panel.
The remnant of SN~1885 (S~And) appears as a small ($0.8''$) round dark spot
of \ion{Ca}{2}~H~\&~K absorption $16''$ southwest of the nucleus.
The bottom panel shows an enlarged section of this same image centered on the SN 1885 region
but shown in a negative log stretch. }
        \label{fig:introfigs}
\end{figure*}

\begin{figure*}[tb!]
    \begin{minipage}{175mm}
    \begin{center}
    \leavevmode
    \includegraphics[scale=.60]{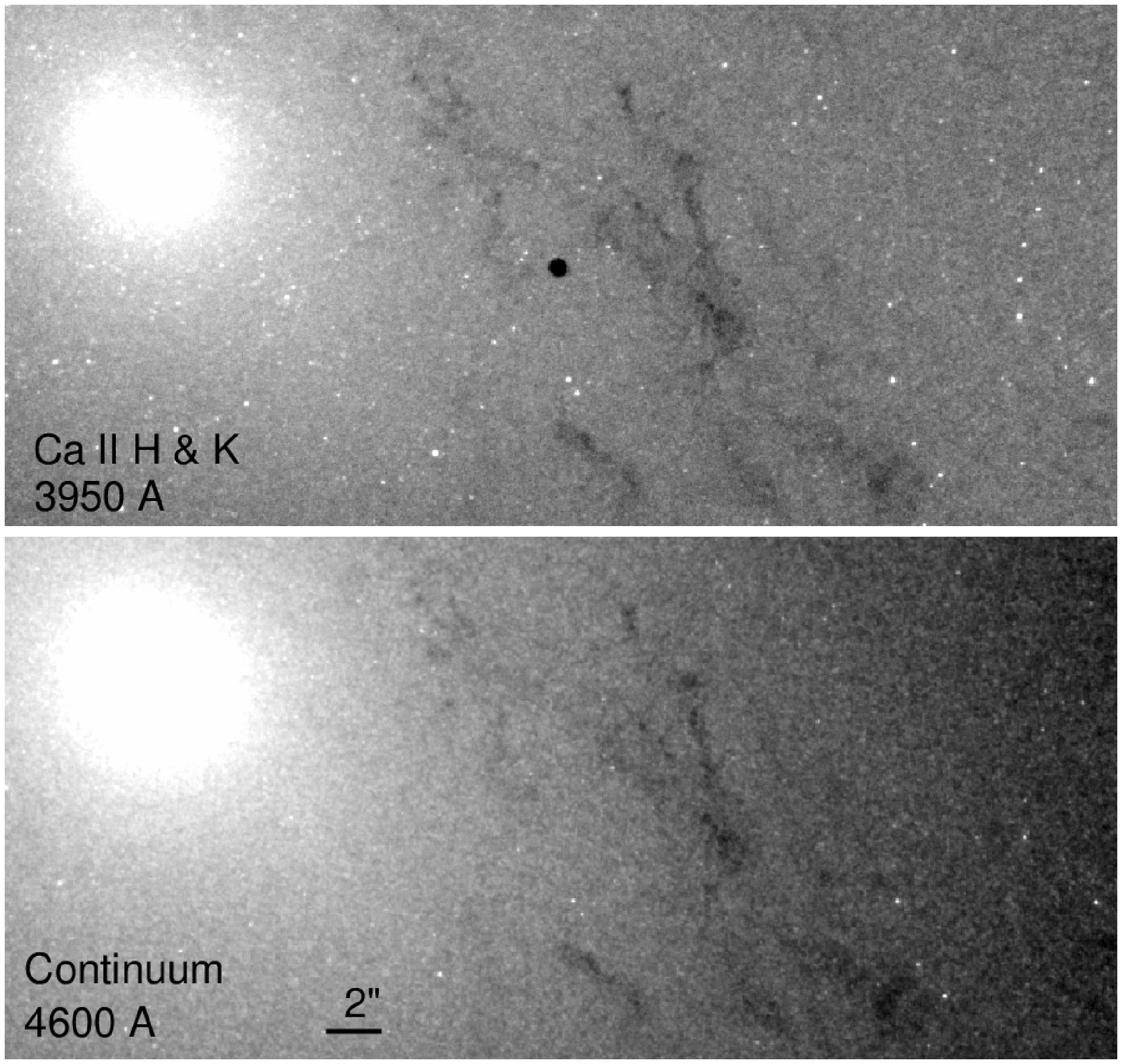}
    \end{center}
    \caption{
    \label{NYSINYD}
{\it{Top:}} ACS/WFC FR388N image of the M31 bulge
shows SNR~1885 prominently visible via 
\ion{Ca}{2} $3934,3968 \unit{\AA}$~K~\&~H absorption.
{\it{Bottom:}} ACS/WFC F462N continuum filter image of the same region
showing no hint of the remnant.
    }
    \end{minipage}
 \end{figure*}

\section{Observations}

Table 1 lists the {\sl HST} images taken or examined as part of this study
of the remnant of SN~1885, hereafter referred to as SNR~1885. These images were
scaled and co-aligned using SAOimage and IRAF image
routines applied to M31 bulge stars detected near the remnant.

The images were obtained using either the ACS Wide Field Channel (ACS/WFC),
the ACS High Resolution Channel (ACS/HRC), or the WFC3/UVIS camera.  The
ACS/WFC detector consists of two $2048 \times 4096$ CCDs covering a field of
view $202'' \times 202''$ with an average pixel size of $0\farcs05$. The
ACS/HRC consists of a single $1024 \times 1024$ CCD providing a spatial
resolution of $0\farcs025 \times 0\farcs025$ pixel$^{-1}$ and a nominal $29''
\times 26''$ field of view. The WFC3/UVIS detector has two $2051 \times 4096$
devices yielding a field of view of $162'' \times 162''$ with $0\farcs039$
pixels.
 
Figure \ref{specuv} (top panel) shows a model predicted UV-optical absorption
spectrum of SNR~1885 \citep{Fesen1999}.  Vertical lines, with lengths
proportional to oscillator strengths, mark wavelengths of the strongest lines.
For simplicity, all ions are assumed to have the same velocity profile as the
observed Ca lines. 

The model spectrum for SNR~1885 shown here differs slightly from that shown in
Fig.~4 of \citet{Fesen1999} in that the spectrum shown here is for a narrow
line of sight through the center of the remnant, whereas the spectrum in
\citet{Fesen1999} was convolved through the $0\farcs43$ aperture of the FOS.
Consequently the absorption lines here are slightly broader and deeper.

The strengths of \ion{Ca}{2}, \ion{Ca}{1}, and \ion{Fe}{1} absorptions are from
a fit to the FOS spectrum, while abundances of other elements are from the
delayed detonation model of \cite{HWT98}.  All ions are assumed neutral or
singly-ionized, with the ratio of neutral to singly-ionized being determined by
the lifetime of the neutral against photoionization by UV light from the bulge
of M31 \citep{Fesen1999}.

The middle plot of Figure \ref{specuv} shows the transmission bandpasses for
the WFC3/UVIS F225W,  F275W, F336W, and  F390M filters, along with the ACS/HRC
F250W and ramp FR388N filters.  The filter peak transmissions have been
normalized to unity but do not reflect differences in total system (telescope + camera +
filter) throughputs (for comparisons see \citealt{Larsson13}).

Archival IUE spectrum of the M31 bulge near the location of SNR~1885
\citep{Burstein1988} is shown in the lower panel of Figure \ref{specuv} along
with the {\sl HST}/FOS spectrum of SNR~1885 \citep{Fesen1999}.  The upper line
is the bulge spectrum, while the lower line is the spectrum absorbed through
SNR~1885, expected in the case of IUE, observed in the case of FOS.  M31
bulge's UV flux is relatively weak and drops significantly shortward of 2900
\AA \ but then slowly declines out to 2400 \AA \

\begin{figure*}[t]
        \centering
        \includegraphics[scale=.95]{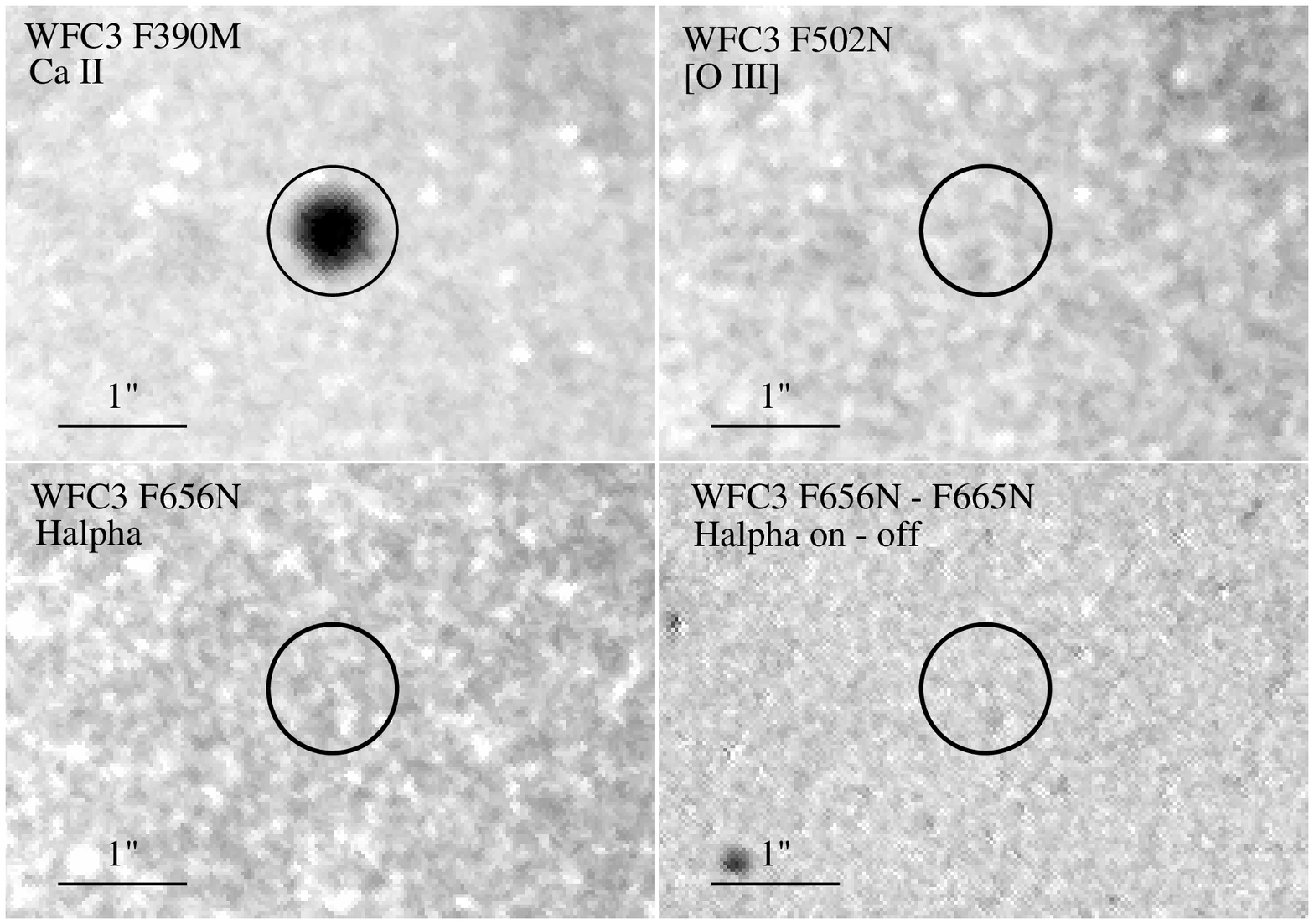}
\caption{ [\ion{O}{3}] and H$\alpha$ images of the site of SN 1885.
Upper left image shows SNR~1885's \ion{Ca}{2} absorption as a reference.
Lower right image is a difference image constructed
from H$\alpha$ ``on'' and ``off'' F656N and F665N images. Circles are 1 arcsecond in diameter.
}
\label{O_n_Halpha_images}
\end{figure*}


\subsection{Previous SNR~1885 and M31 Bulge Images}

Prior ACS/WFC and ACS/HRC images taken of SNR~1885 have been presented and
discussed by \citet{Fesen2007}.  These images included \ion{Ca}{1} (FR423N) and
\ion{Ca}{2} (FR388N) images, along with 4600 \AA \ and UV continuum images
taken with filters FR462N and F330W respectively to allow background subtraction for the
\ion{Ca}{1} and \ion{Ca}{2} images.

These images also included a $15{,}600 \unit{s}$ ACS/HRC exposure with the
F250W filter which resulted in a weak detection of the remnant in absorption.
This filter encompasses the strong \ion{Fe}{2} $2585, 2599 \unit{\AA}$ resonance
lines but also covers the \ion{Mg}{2} $\lambda\lambda2796, 2803 \unit{\AA}$
doublet and \ion{Mg}{1} $2852 \unit{\AA}$ leaving ambiguity as to
how much of the observed absorption should be attributed to \ion{Fe}{2} or to
\ion{Mg}{1} and \ion{Mg}{2}.  The new WFC3 \ion{Fe}{2} image reported in the
present paper uses a shorter wavelength filter, namely F225W, to avoid
contaminated from these lighter elements.

Because M31 has been a target for several {\sl HST} imaging programs,
additional images in the {\sl HST} archive  were examined.  These included 2010
and 2011 images taken as part of an emission line survey of the M31 nucleus
encompassing the SNR~1885 site obtained with WFPC3/UVIS (GO:12174, PI: Z.\ Li).
This survey comprised a set of $2700 \unit{s}$ exposures in H$\alpha$ (F656N),
H$\alpha$ off-band F665N), [\ion{N}{2}] (F658N), [\ion{O}{3}] (F502N), V band
(F547M), \ion{Ca}{2}~H~\&~K (F390M), and [\ion{O}{2}] (F373N).  The \ion{Ca}{2}
F390M passband has a FWHM $\approx 200 \unit{\AA}$ centered at $3900
\unit{\AA}$, about twice the width of the \ion{Ca}{2} FR388N ramp filter which
was centered at $3950 \unit{\AA}$ (see \citealt{Fesen2007}).  No H$\alpha$ or
[\ion{O}{3}] emissions were detected at the location of SNR~1885 in the M31
survey image data (see \S3.1) but absorption was detected in the \ion{Ca}{2}
filter F390M, and also in the nominal [\ion{O}{2}] $\lambda\lambda$3726, 3729
filter (F373N), which can be attributed to \ion{Fe}{1} $3720 \unit{\AA}$
absorption (see below).

\subsection{Fe I Image Data}

Four different filter images detected \ion{Fe}{1} line absorption from
SNR~1885.  An ACS/WFC \ion{Fe}{1} 3720 \AA \ absorption image was obtained during
five orbits on 14 December 2009 using the narrow passband [\ion{O}{2}] ramp
filter FR388N.  The total exposure time was $12{,}700 \unit{s}$, comprising 11
separate images dithered by a four-point square dither pattern.

The ramp filter was positioned so as to yield a peak transmission at the
SNR~1885 site of 45\% for line center of the \ion{Fe}{1} line at $3720
\unit{\AA}$.  The filter's narrow passband (FWHM $\approx 70 \unit{\AA}$;
Fig.~\ref{specuv}) limits detection of \ion{Fe}{1} $3720 \unit{\AA}$ absorption
mainly to expansion velocities of $\pm 3000 \unit{km} \unit{s}^{-1}$, although
the full $100 \unit{\AA}$ bandpass is sensitive to velocities as high as $\pm
5000 \unit{km} \unit{s}^{-1}$.  Because the filter's transmission was only 2\%
at $3763 \unit{\AA}$, it effectively blocked contamination from high-velocity
\ion{Ca}{2} $3934 \unit{\AA}$ absorption (i.e., $3934 \unit{\AA}$ blueshifted
by $13,000$ km s$^{-1}$ is $3763 \unit{\AA}$).

A similar image showing \ion{Fe}{1} absorption from SNR~1885 was obtained in
December 2010 using the WFC3/UVIS filter F373N and taken as part of an emission
line survey of the bulge of M31. The exposure time was 2700 s. The WFC3 F373N
filter has a narrower bandpass (FWHM $\approx 50 \unit{\AA}$) compared to the
ACS/WFC FR388N filter and a redder centroid of $3730 \unit{\AA}$.  This means
that the WFC3 F373N image is sensitive to a slightly different range of
\ion{Fe}{1} 3720 \AA \ velocities relative to the ACS FR388N image, but still
uncontaminated by \ion{Ca}{2} $3934 \unit{\AA}$ absorption.

Two broad passband ACS/HRC and WFC3/UVIS filter images also detected the
remnant's \ion{Fe}{1} absorptions.  An ACS/HRC F330W image with a total exposure
time of 5200 s was taken in October 2004, and a three WFC3 F336W
filter images totaling 5220 s were obtained in July 2010. These ACS/HRC F330W
and WFC3/UVIS F336W filters mainly provide information on an adjacent
wavelength region relatively free of strong absorption features and thus are
are primarily sensitive to the remnant's stellar background.  However, both
filters cover the \ion{Fe}{1} line at 3441 \AA \ 
and thus both show  \ion{Fe}{1} absorption from SNR~1885. 
These images are also weakly sensitive to blueshifted
\ion{Fe}{1} 3021 \AA\ absorption.  While these camera + filters combinations
have similar bandpasses, the integrated system throughput for the WFC3/UVIS
F336W is nearly twice that of the ACS/HRC F330W resulting in a higher S/N.


\begin{figure*}[t]
\begin{minipage}{175mm}
\begin{center}
\leavevmode
\includegraphics[scale=.9]{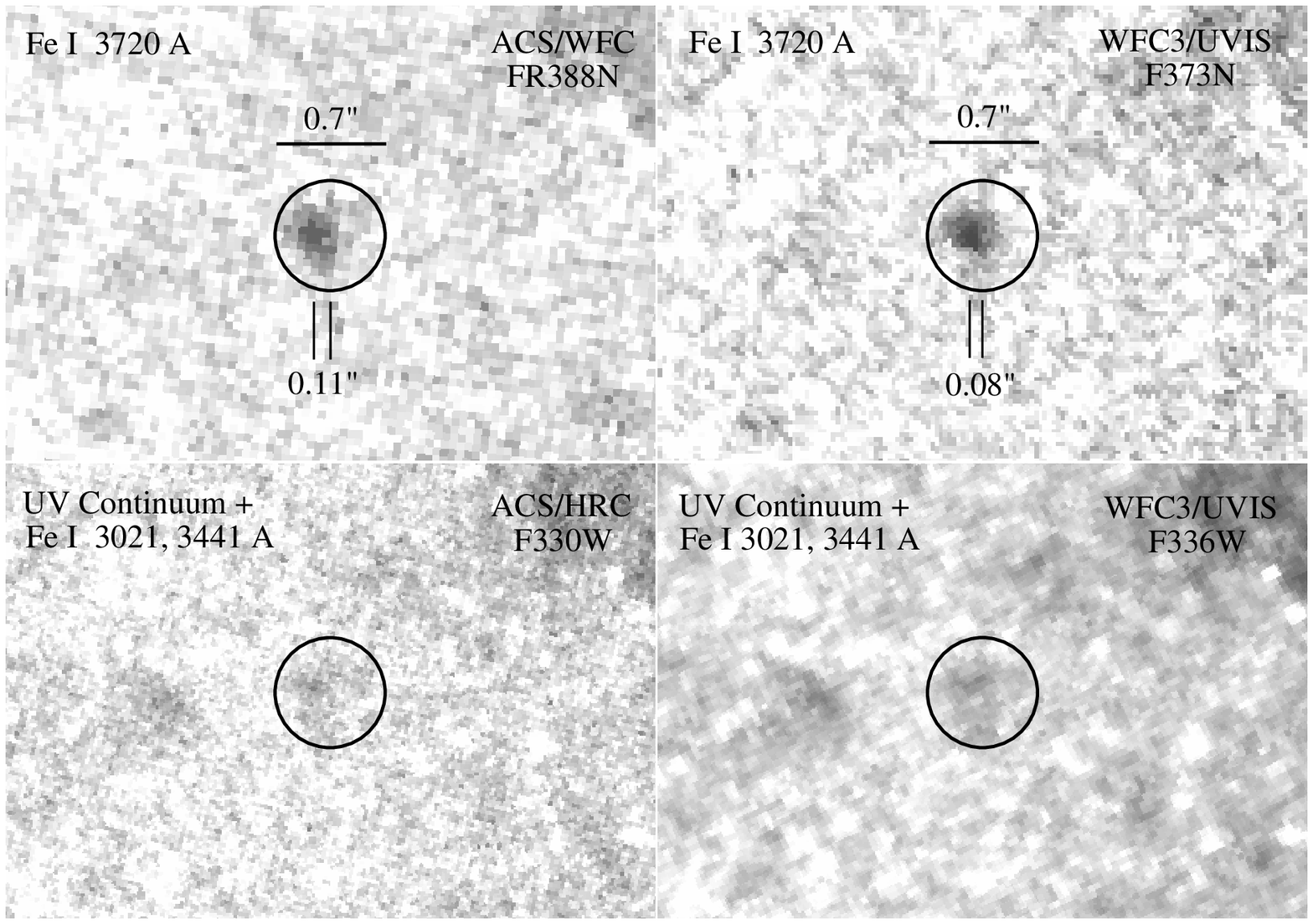}
\end{center}
\caption{
\label{FeI_images}
{\it{Top Row:}}
ACS/WFC FR388N and WFC3/UVIS F373N images
of SNR~1885, sensitive to \ion{Fe}{1} $3720 \unit{\AA}$ absorption.
{\it{Bottom Row:}}
ACS/HRC F330W and WFC3/UVIS 336W images sensitive
to \ion{Fe}{1} resonance lines at 3020 and 3441 \AA.
and should be relatively uncontaminated by absorption from lighter elements.
The $0\farcs7$ diameter circle on each panel
marks the extent of strong \ion{Ca}{2} absorption.
North is up, east to the left for all images.   }
\end{minipage}
\end{figure*}

\begin{figure*}[t]
        \centering
\includegraphics[scale=.9]{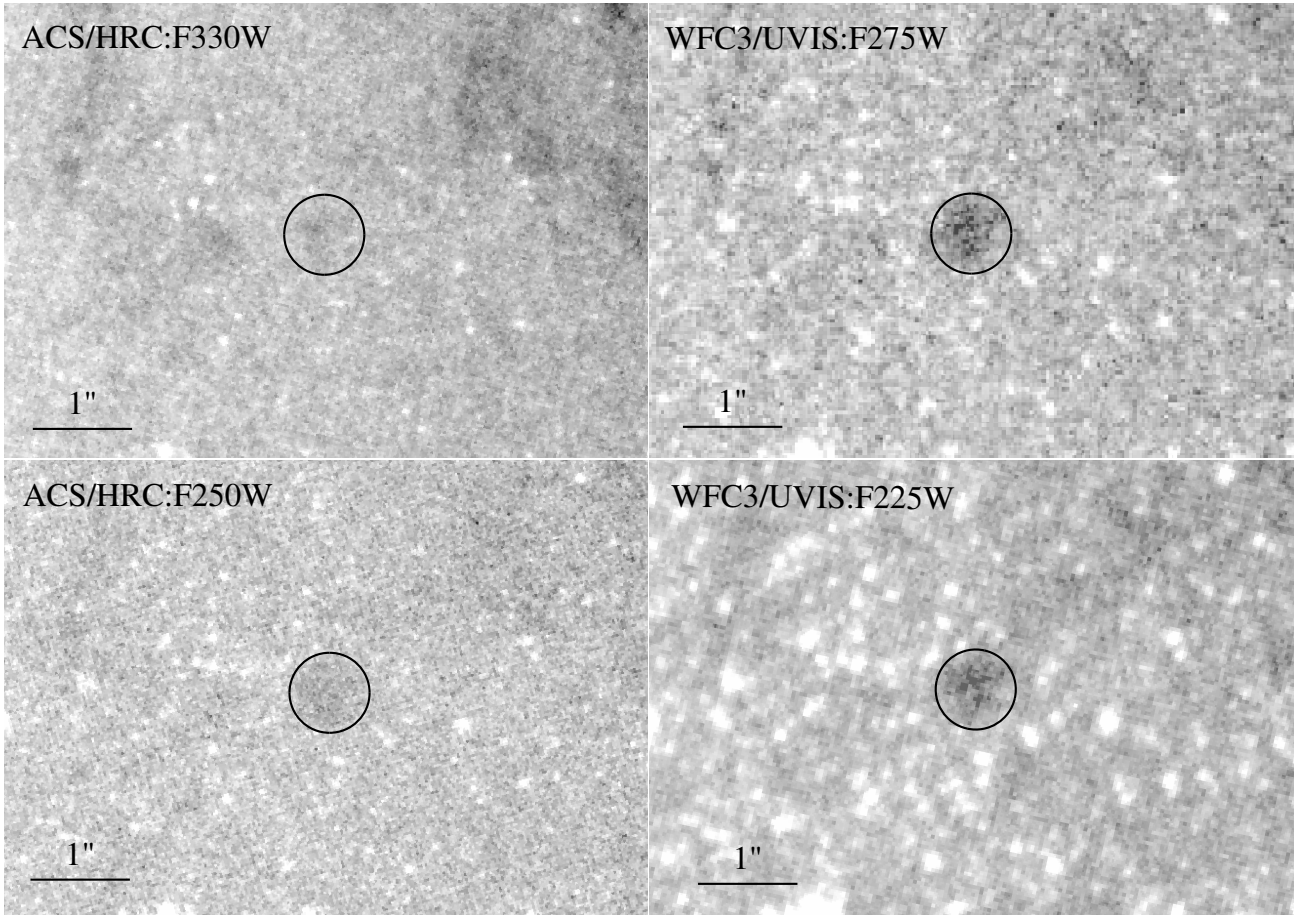}
\caption{Near UV images of SNR~1885.
showing the remnant's detection
Upper panels: ACS/HRC F330W filter image (left)
sensitive to \ion{Fe}{1} and 3100 to 3700 \AA \ continuum,
and a WFC3 F275W filter image (right) sensitive to 2500 to 3050 \AA \ flux.
Lower panels: An ACS/HRC F250W image (left) sensitive to 2400 to 3200 \AA \ flux,
and a WFC3 F225W filter image (right) sensitive to 2100 to 2700 \AA \ flux.  }
        \label{UV_images}
\end{figure*}


\begin{figure*}[t]
 \includegraphics[scale=.95]{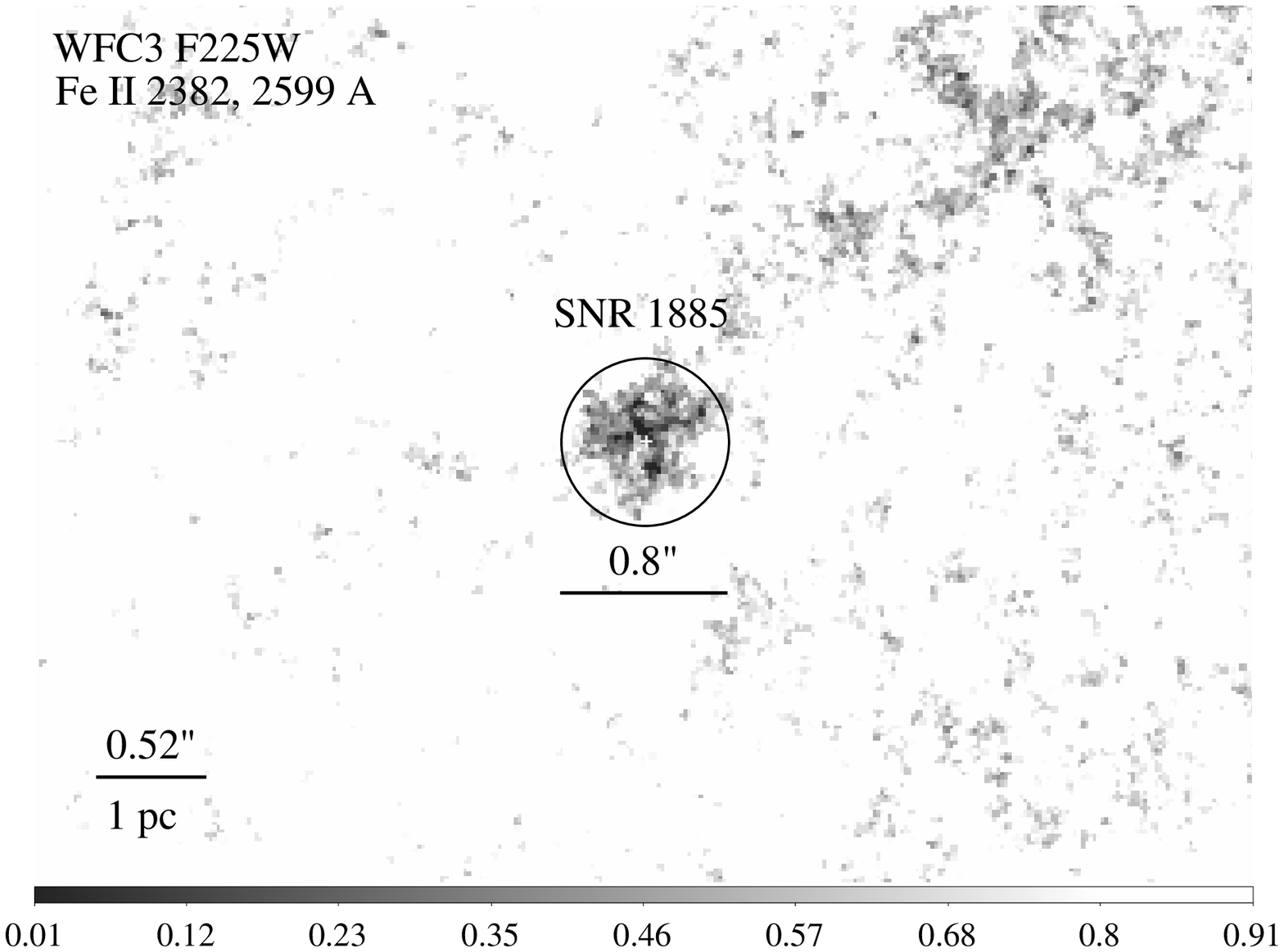}
 \caption{The WFPC3 F225W \ion{Fe}{2} image of SNR~1885 shown in a high contrast, linear intensity stretch. 
The circle (radius of $0\farcs4$) and its center (white cross) indicates the maximum  
extent and center of the remnant's detected \ion{Ca}{2} H \& K line absorption.
Expansion of $0\farcs4$ over a 125 yr time span corresponds to an average expansion velocity
$\simeq$ 12,000 km s$^{-1}$ at M31's 785 kpc distance. } 
  \label{Fe2_closeup}
\end{figure*}


\subsection{Fe II Image Data}

Imaging the remnant's \ion{Fe}{2} required using filters sensitive to the
wavelength region between 2300 and 2600 \AA.  Bandpasses for the WFC3 F225W,
F275W, and F336W filters are shown as shaded regions in Figure \ref{specuv}.

The WFC3/UVIS F225W filter is the best for detecting SNR~1885's \ion{Fe}{2}
absorption free from contamination from other sources.  It has a FWHM $\approx
500 \unit{\AA}$ bandpass with a peak transmission of 8.6\% centered around
$2300 \unit{\AA}$, as illustrated in Figure~\ref{specuv}.  The principal expected
absorbers in this band are the strong UV resonance lines of \ion{Fe}{2} at
$2343, 2382 \unit{\AA}$, with small contributions from \ion{Fe}{1} $2483,
2523 \unit{\AA}$ and \ion{Ni}{1} around $2300 \unit{\AA}$.  The filter also
catches the strong \ion{Fe}{2} lines at $2586, 2599 \unit{\AA}$ in its red
wing.

Thus this filter's bandpass provides detection dominated by \ion{Fe}{2} and
largely free of contamination by lighter elements.  In contrast, both the WFC3
F275W and the ACS/HRC F250W filter have bandpasses sensitive to \ion{Mg}{1} and
\ion{Mg}{2} in addition to \ion{Fe}{2} (Fig.\ \ref{specuv}).

A series of WFC3 F225W images of SNR~1885 were obtained during nine orbits split into
three visits that occurred on 17 and 21 November and 1 December 2012.  The
resolution of the WFC3 camera is $0\farcs039$ per pixel giving it a slightly
better angular resolution scale than the $0.049''$ of the ACS/WFC \ion{Fe}{1}
image and previous WFPC2 images of \ion{Ca}{1} and \ion{Ca}{2}
\citep{Fesen1999}.  

Twelve individual images were taken during each of the three visits, dithered
with a primary three-point and secondary four-point pattern. There was a small
error in guiding for the third set of images (ibr803030) which resulted in a
slightly smeared image in one direction.  The images from the three visits were
co-aligned with IRAF tasks and stacked into one final $3 \times 8064
\unit{s} = 24{,}192 \unit{s}$ exposure.

\begin{figure*}[t]
        \centering
        \includegraphics[scale=.95]{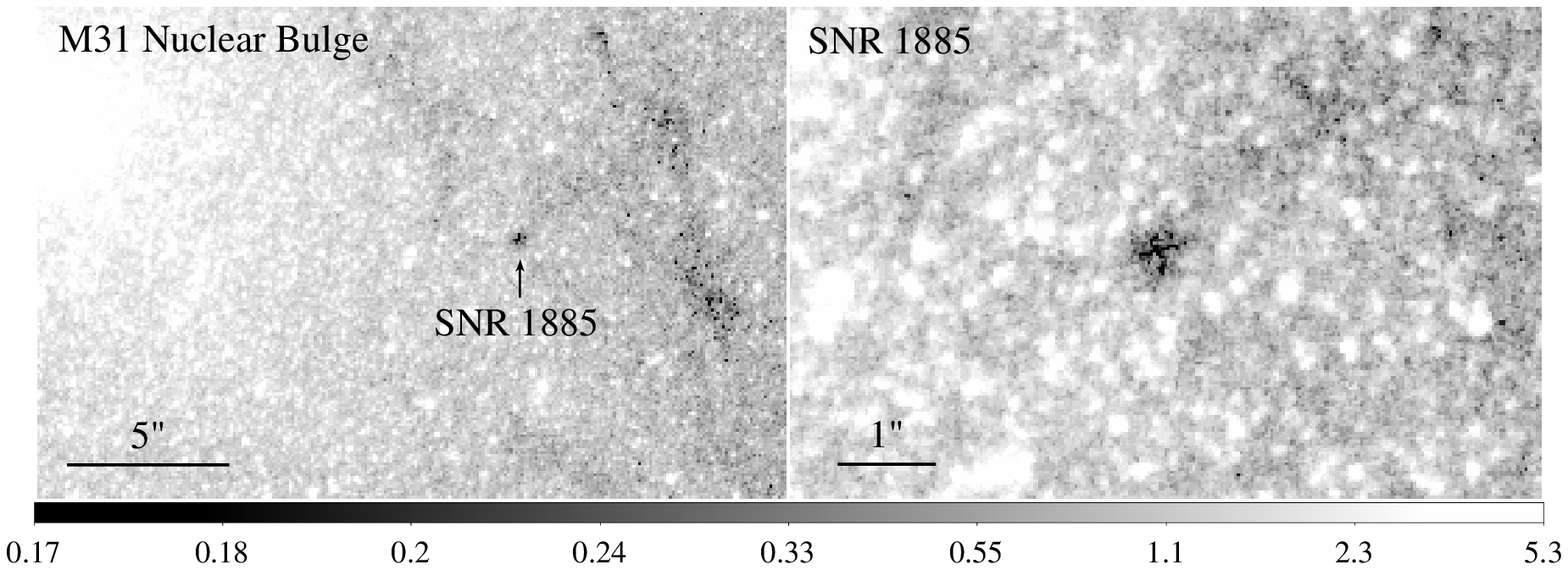}
        \includegraphics[scale=.95]{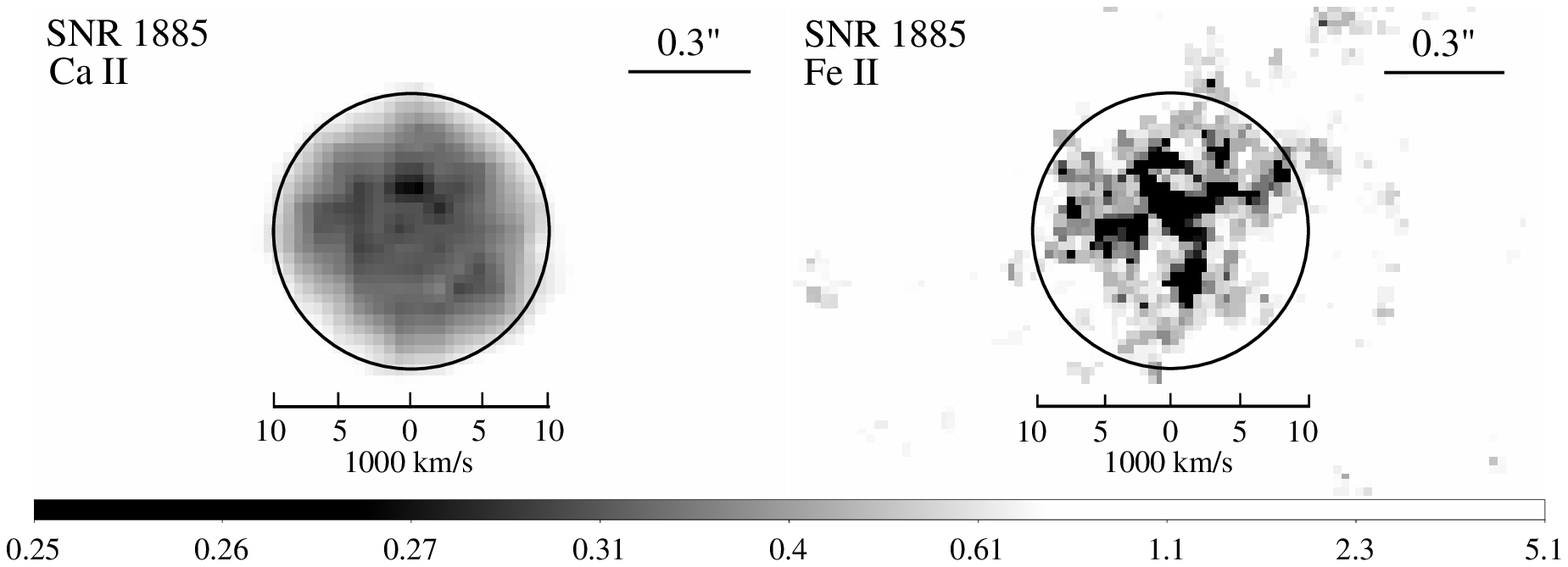}
\caption{Upper panels: Wide and narrow WFC3 F225W views of M31's western bulge region
showing SNR~1885 in context with neighboring dust lanes. 
Lower panels: Log intensity stretched views of SNR~1885 as seen in \ion{Ca}{2} and \ion{Fe}{2} images.
The circles shown (radius = $0\farcs342$) correspond to an average expansion velocity of 10,000 km s$^{-1}$ since 1885.
Note the presence of clumpy, interior
\ion{Ca}{2} absorption and the extent of \ion{Fe}{2} absorption  ``plumes''out past these \ion{Ca}{2} absorption clumps
reaching in places to the edge of the
\ion{Ca}{2} absorption.   
The intensity bar refers to both the \ion{Ca}{2} image and \ion{Fe}{2} images. 
}
\label{Ca_Fe_images}
\end{figure*}


One additional orbit was used to obtain an off-band image of the remnant and
M31 bulge. The off-band image used the WFC3 F336W filter which has a FWHM
$\approx 550 \unit{\AA}$ bandpass with a peak transmission of 19\% around $3300
\unit{\AA}$, as illustrated in Figure~\ref{specuv}.  The total exposure
of this dithered image was $2724 \unit{s}$. Archival WFC3 F336W images of were
also examined.  


\section{Results}

Ideally one would like images of SNR~1885 to be spatially well-resolved and of
high signal-to-noise.  However, the small size ($0\farcs8$ diameter) and
faintness of the bulge of M31 in the ultraviolet make such images challenging
to secure.  Moreover, interpretation can be ambiguous in wavelength bands where
absorptions from multiple species and ionization states overlap.

From this perspective, the cleanest images are the high resolution \ion{Ca}{2}
FR338N image reported by \citet{Fesen2007} and the \ion{Fe}{1} and
\ion{Fe}{2} images presented here.  Both \ion{Ca}{2} and \ion{Fe}{2} are
expected to be the dominant ion of their element --- the singly-ionized ion ---
and both species are expected to dominate absorption in the respective filters.
As will be discussed below, the \ion{Ca}{2},
\ion{Fe}{1}, and \ion{Fe}{2} images reveal distinctively different structures.

\subsection{\ion{Ca}{2} in SNR~1885}

Figure \ref{fig:introfigs} shows a 2700 s {\sl HST} WFC3 F390M filter image of
the central region of M31. SNR~1885 is the small circular spot (dark in the
upper panel and white in the lower panel) approximately $10''$ west and $4''$
south of the M31's nucleus.  Taken using the 200 \AA \ wide F390M filter, the
presence of SNR~1885 is striking against M31's bulge light
and places the remnant in context within the galaxy's nuclear region. 

The remnant is visible in \ion{Ca}{2} H \& K line absorptions from the
remnant's Ca-rich ejecta.  Images such as this illustrate how prominently the
remnant of SN~1885 can appear in absorption.

SN 1885 fortunately exploded on the near side of the bulge, so that its
expanding remnant is well back-lit by M31 bulge stars.  It is also a fortuitous
that the remnant's projected location lies some distance away from several M31
bulge dust lanes (see Fig.\ \ref{fig:introfigs}) thus making analysis of the
remnant's absorption less confused.

Although seen prominently in filter images sensitive to strong resonance lines,
such as \ion{Ca}{2}, SNR~1885 is virtually invisible at nearby continuum
wavelengths.  Figure \ref{NYSINYD}, reproduced from \citet{Fesen2007},
demonstrates how SNR~1885 can be so clearly detected in filter images that are
sensitive to strong resonance lines, such as in this case the \ion{Ca}{2} 3934,
3968 \AA \ H \& K lines, whereas it is virtually invisible at neighboring
wavelengths, in this case 4600 \AA \ using the F462N filter which lack any
significant resonance line features. The lack any any hint of the remnant in
the 4600 \AA \ image also indicates that SN 1885 generated virtually no
appreciable dust. This situation along with the lack of any detectable
H$\alpha$ or [\ion{O}{3}] line emissions associated with the remnant (see Fig.\
\ref{O_n_Halpha_images}) helps explain the lack of any optical detection of
SNR~1885 before its discovery by \citet{Fesen89} despite several 
searches.

\subsection{\ion{Fe}{1} in SNR~1885}

Figure \ref{FeI_images} shows four {\sl HST} ACS and WFC3 images of SNR~1885
which are sensitive to the remnant's near-UV \ion{Fe}{1} absorption lines. In
order to make clear the location of the remnant, a $0\farcs7$ diameter circle
is shown on these images to mark the extent of the remnant's prominent
\ion{Ca}{2} H \& K absorption.

The upper left panel shows an ACS/WFC \ion{Fe}{1} 3720 \AA \ absorption image.  A
similar image is shown in the upper right hand panel which shows a F373N
exposure obtained with WFC3/UVIS. The \ion{Fe}{1} absorption is seen extended
in both images with a diameter of $\approx 0\farcs4$ and displaced eastward
from the remnant's center as defined by  \ion{Ca}{2} images.

While these images are fairly similar in appearance, there are slight
differences.  The peak of \ion{Fe}{1} absorption is displaced eastward
$0\farcs11 \pm 0\farcs03$ from remnant center in the FR388N image but slightly
less so ($0\farcs08 \pm 0\farcs03$) in the F373N image. Such a difference is
only marginally significant given the S/N of the data but, if real, may simply
be due to differences in the two filter bandpass widths and centers.  Filter
bandpass differences probably also account for differences seen in the
darkest (most absorbed) regions of SNR~1885's due to \ion{Fe}{1} 3720 \AA;
$0.60 \pm 0.05$ and $0.40 \pm 0.07$ times the bulge background flux for 
the FR388N and F373N images, respectively.

More importantly, both images show an \ion{Fe}{1} absorption patch that is
distinctly smaller than that of either \ion{Ca}{2} or \ion{Ca}{1}
\citep{Fesen2007}.  Also, unlike the remnant's \ion{Ca}{2} or \ion{Ca}{1}
absorptions, the detected \ion{Fe}{1} absorption gives no sign of being
confined to a shell. 

Broader passband ACS/HRC F330W and WFC3/UVIS F336W images of the remnant are
shown in the lower two panels of Figure \ref{FeI_images}.  Both images show
weak absorption within the remnant (as marked by the circle indicating the
extent of strong \ion{Ca}{2} absorption) due mainly to \ion{Fe}{1}
3441 \AA \ absorption. The amount of light bulge light blocked by SN~1885 in
the ACS F330W and WFC3 F336W images potentially due to both \ion{Fe}{1} 3021
and 3441 \AA \ line absorptions within the $0\farcs7$ circles shown in Figure 4
is only $\simeq$5\%, indicating that these images are mainly continuum images.

These images again show a eastward displacement of the peak \ion{Fe}{1}
absorption.  However, unlike the \ion{Fe}{1} images taken with narrow passband
filters seen in the upper panels, these images reveal a larger patch of
\ion{Fe}{1} absorption, extending farther to the south and west, with the
absorption appearing less concentrated than in the narrow filter
images. The full area of \ion{Fe}{1} absorption is best seen in the WFC3 F336W
image which has higher S/N due to greater camera + filter throughput.

While the observed off-center displacement of \ion{Fe}{1} mass could suggest a
velocity asymmetry around 3500 km s$^{-1}$, similar to the maximum values
obtained for SNe~Ia from late-time, near-IR spectra \citep{Maeda2010}, there is
reason to doubt this explanation given the similar eastern offset of the
remnant's \ion{Ca}{1} \citep{Fesen2007}. \ion{Fe}{1}, like \ion{Ca}{1}, has a
relatively short lifetime in SNR~1885, only about $10 \unit{yr}$ due to
photoionization by UV light from the bulge of M31. 

The persistence of \ion{Fe}{1} and \ion{Ca}{1} some 130 years after explosion
suggests that these ions may be self-shielded by their own continuum optical
depth \citep{HF91,Fesen2007}. The current presence of \ion{Fe}{1} and
\ion{Ca}{1} concentrated along the eastern half of the remnant and
unexpectedly {\it{toward}} the M31 nucleus, suggests an important source of
ionizing UV photons along the remnant's projected western hemisphere. If this
is the case, the remnant's Fe and Ca rich ejecta in its western half could have
self-shielded ejecta in the remnant's central and eastern regions leading to
the different \ion{Ca}{1}/\ion{Ca}{2} and \ion{Fe}{1}/\ion{Fe}{2} absorption
distributions observed. Since \ion{Fe}{1} likely represents only a small
percentage of the remnant's Fe-rich ejecta mass, \ion{Fe}{1} images may not
accurately portray the true distribution of SNR~1885's Fe-rich material.
 
\subsection{\ion{Fe}{2} in SNR~1885}

The three {\sl HST} images taken of SNR~1885 which cover the near-UV
wavelengths of strong \ion{Fe}{2} resonance lines \citep{Morton91} are shown in
Figure \ref{UV_images}. This figure also includes the broad passband ACS/HRC
F330W discussed above in regard to the remnant's \ion{Fe}{1} absorption and is
included here to show the stellar background and dust lanes at and around the SNR~1885
site.  All four images are shown using a square root intensity stretch in order
to show both the remnant's absorption features and neighboring bulge features. 

These images show numerous bright bulge stars in the vicinity of SNR~1885
which are probably mainly red giants with strong UV coronal line emissions.
Individual bulge stars become increasingly prominent at shorter wavelengths
and are most apparent in the F225W filter image seen in the lower right
panel. The F330W image, with a passband comparatively free of line absorptions
suggests that with the exception of two stars along the remnant's
westernmost limb, no bright foreground or background stars lie in the direction
to SNR~1885. 

Although the F225W, F250W, and F275W filter passbands cover some or all of the 
strong \ion{Fe}{2} lines at $2343, 2382, 2586$ and 2599 $\unit{\AA}$, the detected
absorption in SNR~1885 appears different in each filter. Below, we discuss 
possible reasons for these differences.
 
We begin with the ACS/HRC F250W filter image which has been presented and
discussed by \cite{Fesen2007}. This image yielded the weakest detection of the
remnant.  Given the HRC's small pixels and only a 6\% integrated system
throughput for the ACS/HRC + F250W filter, the resulting low S/N image showed a
featureless absorption spot but one whose diameter is nearly the same as that
of the remnant's \ion{Ca}{2} absorption.  The F250W filter has a relatively
broad passband which is starts at 2300 \AA \, peaks at 2500 \AA \ and extends
to 3300 \AA. This makes this filter sensitive to several strong resonance lines
including \ion{Fe}{1}, \ion{Fe}{2}, \ion{Mg}{1}, and \ion{Mg}{2}. Thus, this
image is likely a composite of Fe and Mg absorption features, limiting its
usefulness for determining the remnant's \ion{Fe}{2} 2D distribution.

The WFC3 F275W image is fairly similar to the F250W image but with hints of
some internal structure.  Like that of
the F250W filter, F275W's bandpass encompassed many likely absorption lines
including \ion{Fe}{1}, \ion{Fe}{2}, \ion{Mg}{1}, and \ion{Mg}{2}.

The most plausible explanation of differences between the F250W and F275W
images is that the F275W image is picking up substantial more absorption from
resonance lines of \ion{Mg}{2} $2796, 2804 \unit{\AA}$ and \ion{Mg}{1} $2853
\unit{\AA}$ due to the WFC3's higher throughput at 2800 \AA. Note that the
model spectrum shown in the top panel of Figure~\ref{specuv} is for a line of
sight directly through the center of SNR~1885.  If SNR~1885 has Fe concentrated
to the center and lighter elements concentrated in a shell, then a line of
sight through the outer parts of SNR~1885 would show relatively more absorption
from Mg and less from Fe.

The F225W WFC3 image (lower right panel in Figure \ref{UV_images}) 
provides the cleanest image of Fe II absorption in SNR~1885,
uncontaminated by Mg absorption features.
As illustrated in Figure ~\ref{specuv}, the
strongest absorption lines in the WFC3 F275W filter are \ion{Fe}{2} $2586, 2599
\unit{\AA}$, while the strongest lines in the WFC3 F225W filter are \ion{Fe}{2}
$2343, 2382 \unit{\AA}$.  Since the strongest lines in both filters are those
of \ion{Fe}{2}, we had expected that the two images would show essentially the
same structure.  But this is not so: the F275W image shows absorption that is
smoother, more extended, and less centrally concentrated than F225W
likely due to added \ion{Mg}{1} and \ion{Mg}{2} absorption.

The WFC3 F225W \ion{Fe}{2} image shows
a structure that differs strikingly from that of either Ca or \ion{Fe}{1}.  It reveals 
four prominent \ion{Fe}{2} absorption streams or plumes
that extend from the center of the remnant to near the outer extent of the
\ion{Ca}{2} absorption.  

Figure~\ref{Fe2_closeup} shows an enlargement of the \ion{Fe}{2} F225W image.
The four apparent \ion{Fe}{2} streams appear clumpy, have similar thicknesses
($\sim 0\farcs1$), and intersect at a point  $\simeq 0\farcs05$ ($\simeq$ 1500
km s$^{-1}$) north of the center of the remnant as defined by the remnant's
\ion{Ca}{2} absorption.  While the detected \ion{Fe}{2} absorption is strongest
in these four plumes, there is also considerable absorption outside these
finger-like plumes extending out to a radius $\simeq$ $0\farcs3$ ($\simeq$
10,000 km s$^{-1}$).  

A major concern in interpreting the F225W \ion{Fe}{2} image is uncertainty
regarding the distribution of background stellar sources as they could affect
the distribution of observed light and dark (absorption) features.  After all,
the M31 bulge around SNR~1885 exhibits a high density of strong stellar sources
in the F225W image.  For example, a few bright UV stars located in front of or
behind the remnant could potentially lead to an uneven absorption structure not
unlike that seen in the F225W image.

Although we cannot rule out this possibility, we view this as unlikely. From
the predicted UV spectrum of SNR~1885 shown in Figure \ref{specuv}, it is not
possible to have a clean ``off-band'' \ion{Fe}{2} image around 2500 \AA \ from
which one could determine the number and brightness of background stars toward
SNR~1885.  However, as noted above, the ACS/HRC F330W image indicates no bright
line-of-sight stellar contamination, especially toward the remnant's center.  

Moreover, while the density of stars in the F225W image surrounding SNR~1885 is
fairly high ($\sim$ 5--8 stars arcsec$^{-1}$, there are several immediately
adjacent bulge regions just to the west $\sim 1''$ in diameter largely free of
bright stars.  Further, the nearly orthogonal plume structure observed in the
F225W image is unlikely to be due to a chance distribution of background stars.
If the similar thickness \ion{Fe}{2} plumes intersecting near at the remnant's
center and occupying only the inner 2/3rds of the remnant's radius was caused
by a chance arrangement of bright background stars, then their presence would
be expected to have a significant impact on the otherwise smooth and largely
spherical \ion{Fe}{2} absorption detected in between the plumes.
 
In addition, nothing so striking in morphology as seen in the F225W image is
visible in any neighboring M31 bulge dust lanes.  This is shown in the upper
panels of Figure \ref{Ca_Fe_images} where we present a wider and log intensity
stretched view of the F225W image.  The depth of the apparent \ion{Fe}{2}
absorption in SNR~1885 seen in the F225W image is equal to or greater than that
seen in neighboring M31 bulge dust lanes. 

The darkest (most absorbed) regions in the F225W image of SNR~1885 are less
than 0.15 times the diffuse (no stars) bulge background flux, with the
remnant's inner $0\farcs4$ diameter region encompassing all four plumes is
$\leq$ 50\% below that of the least bright adjacent bulge regions.  The total
measured flux in the F225W filter for SNR~1885 within a circular region shown
in Figure \ref{Ca_Fe_images} (r = $0\farcs34$) is $0.60 ^{+0.50}_{-0.15} $ that of neighboring
bulge regions at similar distances from the M31 nucleus. 

The four \ion{Fe}{2} plumes exhibit different projected lengths, with
the one to the west extending the farthest, nearly out to the edge of the
remnant's strong \ion{Ca}{2} absorption shell ($0\farcs34$; see Fig.\
\ref{Ca_Fe_images}) implying a maximum transverse velocity $\simeq$ 10,000 km
s$^{-1}$.  If these \ion{Fe}{2} plumes truly trace the distribution of the
remnant's Fe-rich ejecta, the different lengths could arise from line of sight
projections of nearly equal length plumes but expanding in different
directions.  

SNR~1885's 2D \ion{Fe}{2} absorption stands in stark contrast to its
\ion{Ca}{2} absorption distribution, also shown in Figure \ref{Ca_Fe_images}
for easy comparison.  Whereas the remnant's \ion{Ca}{2} is concentrated in a
clumpy shell spanning a range 1000 to 5000 km s$^{-1}$ in expansion velocity,
the remnant's Fe-rich ejecta is concentrated in the four finger-like plumes
which extend well past the Ca-rich shell.  Of equal importance is the apparent
presence of Fe-rich material out to $\approx$10,000 km s$^{-1}$ in transverse
velocity in non-plume directions.

The significance of the apparent $\simeq$ 1500 km s$^{-1}$ ($0\farcs05$) off-center
intersection of the \ion{Fe}{2} absorption fingers relative to the remnant's
\ion{Ca}{2} absorption is difficult to assess given indications that the
remnant's \ion{Ca}{2} absorption is not perfectly symmetric.  The maximum
\ion{Ca}{2} absorption seen in both ACS FR388N and WFC3 F390W images exhibits a
slight extension ($\sim 0\farcs07$ off the southwestern limb; see Fig.\ 5 in
\citealt{Fesen2007}). Moreover, the remnant's $0\farcs80$ \ion{Ca}{2} diameter seen
in the most recent 2010 images indicates a somewhat smaller 11,900 $\pm 1500$
km s$^{-1}$ transverse velocity than the 12,400 km s$^{-1}$ estimated by
\citet{Fesen2007} and lower but still within the uncertainty of the 13,100
$\pm 1500$ km s$^{-1}$ radial velocity measured from the remnant's \ion{Ca}{2}
H \& K absorption profile \citep{Fesen1999}. Small radial and transverse
velocity differences might indicate a slight non-spherical expansion, meaning
an apparent off-center displacement of the \ion{Fe}{2} absorption may not be
significant and not a clue for an off-center ignition point of SN~1885.  

\section{Discussion}
\label{implications-sec}

We will now discuss how the new {\sl HST} observations of the remnant
of S Andromeda (SNR~1885) can be placed in context of our current
understanding of Type Ia supernovae.  
The most informative {\sl HST} images of SNR 1885 presented above are those in
\ion{Ca}{2} absorption and \ion{Fe}{2} absorption (Fig.\
\ref{Ca_Fe_images}). This is because both \ion{Ca}{2} and \ion{Fe}{2} are
expected to be the dominant ion of that element so their absorption reliably
traces the element as a whole, and because the absorption in the chosen
filters, FR338N and F225W respectively, is dominated by the designated ion,
relatively uncontaminated by absorption from other species.

The presence of Fe at the center of SNR~1885 with Ca-rich debris in a shell
around the center is consistent with SN~1885 being a Type~Ia supernova, albeit
likely a subluminous SN~Ia.  Because its debris are still in near free
expansion after 125 years, its retains the density distribution established
shortly after the explosion.  Thus, the distribution of elements in SNR~1885
can provide valuable kinematic information about the general properties and
character of a Type~Ia explosion.

Consistent with evidence from SNe~Ia light curves, the remnant of SN~1885 shows
evidence for both deflagration and detonation.  A characteristic signature of
detonations is an approximately spherically symmetric layered structure of
burned elements.  Hence,  the shell-like appearance of \ion{Ca}{2}, an
incompletely burned element, points to a detonation explosion. {\sl HST} images show a
layered structure for the intermediate mass elements such as Ca in SNR~1885
\citep{Fesen2007}; specifically, the presence of inner clumps of Ca at
velocities between 1,000 and 5,000 km s$^{-1}$ along with Ca debris with
expansion velocity out to $\simeq$12,500 km s$^{-1}$ and thus is strongly suggestive of
a detonation phase.
 
In contrast, the \ion{Fe}{2} F225W image of SNR~1885 shows several 
finger-like plumes that extend from the center out to an expansion velocity of about
$10{,}000 \unit{km} \unit{s}^{-1}$, near the outer edge of the \ion{Ca}{2}
absorption. The length scale of the lumps seen in SNR~1885 is consistent with
Rayleigh-Taylor instabilities commonly found in SNe~Ia simulations during the
early deflagration phase. Though the \ion{Fe}{2} image shows just four plumes,
this may be a projection effect when combining a small number of plumes.
In summary, the combined Ca and Fe distributions in the remnant of SN 1885 is in line
with an off-center delayed detonation SN~Ia event.  

For SNe~Ia, several explosion scenarios are possible
including classical delayed detonation, pulsating delayed-detonation models
\citep{khokhlov91,yamaoka92,khokhlov93,hoeflich95,gamezo2005,roepke07}, or
dynamical merger models \citep{benz90,rasio94,Pakmor11,isern11}. For the newly 
recognized subclass, SNe~Iax \citep{li03,li11,foley13}, models suggested include 
deflagration models with  burning of the outer layers of a WD \citep{kromer13}
or pulsating delayed detonation models \citep{Stritzinger2014}.

However, the seeming lack of appreciable Ca-rich material in the
center argues against strong mixing and in favor of the explosion of a
massive WDs.  The fact of limited mixing of the inner region and the constraint
of plumes to the low velocities seen compared to the mixing predicted by
hydro-models \citep{r02,gamezo03,roepke07,seitenzahl13} suggest some suppression
of the Rayleigh-Taylor instability  possibly through strong magnetic fields
\citep{Hoeflich2004,Hoeflich2013,Penney2014,Remming2014}.

We split the following discussion into in the constrains for $M_{Ch}$
mass models or, more precisely, the constraints for the deflagration phase of
burning, arising from dynamical WD mergers and in the framework of SNe~Iax. 
The approximately spherically symmetric distribution of \ion{Ca}{2} in contrast
to \ion{Fe}{2} plumes which extend in different directions argues against an
explosion with a preferred symmetry axis as might result from an anisotropic
merger of two white dwarfs. 

Within the framework of merging WDs, the large angular momentum will impose a
preferred axis of symmetry which should be apparent in our 2D images but is
clearly not the case.  A detonation front is responsible for burning without
time for the formation of chemical clumps and mixing.  Thus, none of the
dynamical models show large scale, irregular structures if the merging occurs
on time scales of the orbiting WDs.

One exception are the type of violent, dynamical mergers in which two WDs collide.
Simulations show that some of the Fe-rich material is squeezed out into a thin,
`cone-like' structure which can extend to high velocities.  Instabilities by
the decay of radioactive $^{56}$Ni over the time may produce clumps in the thin
structure. However, we see several structures without a regular pattern and
thus the distribution of the intermediate element of Ca in SNR~1885 seems to be
inconsistent with this model.

Finally we consider SN~Iax events which have been classified as a new type of
thermonuclear explosions.  Though the prototype,  SN~2002cx was several
magnitudes dimmer than S Andromeda, the recent SN~2014Z extended the range to
$-18$ mag, similar to S Andromeda. What sets SNe~Iax apart from normal SNe~Ia
is that the former does not show a brightness decline relation and spectra
similar to SN~1991T.  Considering the uncertainties in the light curve of
S Andromeda given its location against the bright M31 bulge background, SNe~Iax
may well be compatible.  

The nature of SNe~Iax is currently under debate and may include
deflagration burning and ejection of the outer layers of a WD  leading to a low
density shell \citep{kromer13} and/or pulsating delayed detonation models
\citep{Stritzinger2014}.  The low-ejecta mass of the first class is unlikely
because it in conflict with the long time scale of re-ionization and the
layered structure. However, pulsation delayed-detonation models can not be
excluded because they can have a wide variety of mixing during the pulsation,
and show a layered structure in the outer layers.  An alternative
interpretation to the structure may be various amounts of mixing during the or
multiple pulsations of the WD.

\section{Conclusions}

The remnant of the probable Type~Ia remnant of SN~1885 (S And) in the bulge of M31 is
visible in absorption against the background of bulge stars.  
High resolution {\sl HST} images of the remnant of SN~1885
in M31 show the following:

1) The distribution of \ion{Fe}{1} material appears small ($0\farcs3$), and
offset to the east by approximately $0\farcs1$ from remnant center. While this
displacement might indicate a velocity asymmetry of the remnant's Fe-rich
material like that proposed by \citep{Maeda2010} based on near-IR spectral
observations of SNe~Ia, a similar displacement of the remnant's \ion{Ca}{1}
absorption suggests it is due to an uneven ionizing UV flux local to the
remnant.  

2) A deep UV image sensitive mainly to \ion{Fe}{2} resonance lines at $2343,
2382 \unit{\AA}$ shows an irregular shaped \ion{Fe}{2} absorption structure
peaked in strength at the center of the remnant with four plume-like extensions
of especially strong \ion{Fe}{2} absorption which  weaken and terminate
near the remnant's outer boundary as marked by the remnant's \ion{Ca}{2}
absorption extent.

The key result of this paper has been to present a spatially resolved {\sl HST}
image of the supernova remnant in absorption in the strong resonance doublet of
\ion{Fe}{2} and to compare that image to a previously obtained absorption image
in \ion{Ca}{2}~H~\&~K.  Since \ion{Fe}{2} and \ion{Ca}{2} are expected to be
the dominant ions of their respective elements, the images in these lines trace
the bulk of iron and calcium in SNR~1885.  Since the remnant is still in free
expansion at the present time, some 130~years after the explosion, the
distribution of elements remains that established shortly after the explosion,
thereby giving valuable insight on the details of the explosion process.

Our finding that iron, the dominant outcome of nuclear burning to completion,
shows four plumes that start at the center of the remnant and reach out to a
maximum velocity of $10{,}000 \unit{km} \unit{s}^{-1}$, near the maximum extent
of Ca is especially significant.  Calcium, a partially burned element, is by
contrast concentrated in a broad and lumpy shell spanning 1000 to 
5000 $ \unit{km} \unit{s}^{-1}$ and extending out to a maximum of
$12{,}500 \unit{km} \unit{s}^{-1}$.

The observed distribution of iron and calcium in the remnant of S And is
consistent with delayed detonation models of Type~Ia explosions, in which the
explosion begins as a deflagration at the center of a white dwarf, which later
transitions to a detonation.  The deflagration drives Rayleigh-Taylor unstable
plumes to near the surface of the white dwarf.  This deflagration burning heats
and expands the white dwarf so that when the explosion turns into a detonation,
unburned elements are burned not to completion but rather to intermediate mass
elements, such as the calcium observed in the S And remnant.

Progenitors with $M_{Ch}$ mass and, in particular, off-center delayed
detonation models are favored by observations including our previous studies of
S~And.  However, such models have difficulty when taking into account the
predicted burning instabilities and mixing during the deflagration phase both
with respect to the brightness decline relation, late-time IR spectra and, at
least for subluminous SNe~Ia, in early time and maximum light optical and IR
spectra. 

The structure observed in S And may serve as a benchmark to decipher the
mechanism to partially suppress instabilities. Moreover, the evidence for the
Rayleigh-Taylor instabilities suggested by the S And observations reported here
support the validity of $M_{Ch}$ models which require a deflagration phase and
dis-favor dynamical or violent mergers.

~~ \\

This research was supported by NASA through grants GO-10722 and GO-12609 from
the Space Telescope Science Institute, which is operated by the Association of
Universities for Research in Astronomy, and the NSF through grants  AST-0708855 and
AST-1008962 to PAH.

{}   


\begin{thebibliography}{}
\bibitem[Benz et al.(1990)]{benz90} Benz, W., Cameron, A.~G.~W., 
         Press, W.~H., \& Bowers, R.~L.\ 1990, \apj, 348, 647 
\bibitem[Bloom et al.(2012)]{Bloom2012} Bloom, J.~S., Kasen, D., 
         Shen, K.~J., et al.\ 2012, \apjl, 744, L17 
\bibitem[Branch et al.(1995)]{branch95} Branch, D., Livio, M., 
         Yungelson, L.~R., Boffi, F.~R., \& Baron, E.\ 1995, \pasp, 107, 1019 
\bibitem[Burstein et al.(1988)]{Burstein1988} Burstein, D., Bertola, 
         F., Buson, L.~M., Faber, S.~M., \& Lauer, T.~R.\ 1988, \apj, 328, 440 
\bibitem[Chevalier \& Plait(1988)]{CP88} Chevalier, R.~A., \& Plait, P.~C.\ 1988, \apjl, 331, L109
\bibitem[Colgate \& McKee(1969)]{CK69} Colgate, S.~A., \& McKee, C.\ 1969, \apj, 157, 623 
\bibitem[de Vaucouleurs \& Corwin(1985)]{deV85} de Vaucouleurs, G., \& 
         Corwin, H.~G., Jr.\ 1985, \apj, 295, 287 
\bibitem[Fesen et al.(1999)]{Fesen1999} Fesen, R.~A., Gerardy, C.~L.,
         McLin, K.~M., \& Hamilton, A.~J.~S.\ 1999, \apj, 514, 195
\bibitem[Fesen et al.(1989)]{Fesen89} Fesen, R.~A., Saken, J.~M.,
        \& Hamilton, A.~J.~S.\ 1989, \apjl, 341, L55
\bibitem[Fesen et al.(2007)]{Fesen2007} Fesen, R.~A., H{\"o}flich, P.~A., Hamilton, 
         A.~J.~S., Hammell, M.~C., Gerardy, C.~L., Khokhlov, A.~M., \& 
         Wheeler, J.~C.\ 2007, \apj, 658, 396 
\bibitem[Foley et al.(2013)]{foley13} Foley, R. J., Challis, P., 
         Chornock, R., et al. 2013, \apj, 767, 57
\bibitem[Gamezo et al.(2003)]{gamezo03} Gamezo, V.~N., Khokhlov, 
         A.~M., Oran, E.~S., Chtchelkanova, A.~Y., \& Rosenberg, R.~O.\ 2003, Science, 299, 77 
\bibitem[Gamezo et al.(2004)]{gamezo2004} Gamezo, V.~N., Khokhlov, 
         A.~M., \& Oran, E.~S.\ 2004, Physical Review Letters, 92, 211102 
\bibitem[Gamezo et al.(2005)]{gamezo2005} Gamezo, V.~N., Khokhlov, 
         A.~M., \& Oran, E.~S.\ 2005, \apj, 623, 337 
\bibitem[Garcia-Senz \& Woosley(1995)]{Garcia1995} Garcia-Senz, D., 
         \& Woosley, S.~E.\ 1995, \apj, 454, 895 
\bibitem[Hamilton \& Fesen(1991)]{HF91} Hamilton, A.~J.~S., \& Fesen, R.~A.\ 1991, in Supernovae,
         10th Santa Cruz Summer Workshop in Astronomy and Astrophysics,
         ed.\ S.~E.~Woosley (Berlin: Springer-Verlag), 656
\bibitem[Hamilton \& Fesen(2000)]{Hamilton00} Hamilton, A.~J.~S., \& Fesen, R.~A.\ 2000, 
         \apj, 542, 779
\bibitem[Hillebrandt et al.(2013)]{Hillebrandt13} Hillebrandt, W., Kromer, M., R{\"o}pke, F.~K.,
        \& Ruiter, A.~J.\ 2013, Frontiers of Physics, 8, 116 
\bibitem[Hillebrandt \& Niemeyer(2000)]{HN00} Hillebrandt, W., \& 
         Niemeyer, J.~C.\ 2000, \araa, 38, 191
\bibitem[H{\"o}flich \& Khokhlov(1996)]{hk96} {H{\"o}flich}, P., 
         {Khokhlov}, A. 1996, \apj, 457, 500
\bibitem[H{\"o}flich, Wheeler \& Thielemann(1998)]{HWT98} H{\"o}flich, P., 
         Wheeler, J. C., and Thielemann, F.K. 1998, \apj, 495, 617
\bibitem[H{\"o}flich et al.(2004)]{Hoeflich2004} H{\"o}flich, P.,
         Gerardy, C.~L., Nomoto, K., Motohara, K., Fesen, R.~A., Maeda, K., Ohkubo,
         T., \& Tominaga, N.\ 2004, \apj, 617, 1258
\bibitem[H{\"o}flich et al.(2013)] {Hoeflich2013} H{\"o}flich, P., 
         Dragulin, P., Mitchell, J., Penney, B., Sadler, B., Diamond, T., 
         Gerardy, C. \ 2013, Frontiers of Physics, 8, 144
\bibitem[H{\"o}flich et al.(1995)]{hoeflich95} H{\"o}flich  P., 
         Khokhlov, A., Wheeler  J.C. 1995, \apj, 444, 211
\bibitem[H{\"o}flich \& Stein(2002)]{hs02} H{\"o}flich P., \&  Stein J.\ 2002, \apj, 568, 771
\bibitem[Hofmann et al.(2013)]{Hof13} Hofmann, F., Pietsch, W.,
         Henze, M., et al.\ 2013, \aap, 555, A65
\bibitem[Howell(2011)]{Howell2011} Howell, D.~A.\ 2011, Nature Communications, 2, 350 
\bibitem[Hoyle \& Fowler(1960)]{hf60} Hoyle, F., \& Fowler, W.~A.\ 1960, \apj, 132, 565
\bibitem[Iben \& Tutukov(1984)]{Iben84} Iben, I., Jr., \& Tutukov, A.~V.\ 1984, \apjs, 54, 335
\bibitem[Isern et al.(2011)]{isern11} Isern, J., Hernanz, M.,
         \& Jos{\'e}, J.\ 2011, Lecture Notes in Physics, Berlin Springer Verlag, 812, 233
\bibitem[Kaaret(2002)]{Kaaret02} Kaaret, P.\ 2002, \apj, 578, 114 
\bibitem[Khokhlov(1991)]{khokhlov91} Khokhlov, A.~M.\ 1991, \aap, 245, L25 
\bibitem[Khokhlov et al.(1993)]{khokhlov93} Khokhlov, A., M{\"u}ller, E., \&
         H{\"o}flich,  P.\ 1993, \aap, 270, 223
\bibitem[Khokhlov(1995)]{khokhlov95} Khokhlov, A.~M.\ 1995, \apj, 449, 695 
\bibitem[Kromer et al.(2013)]{kromer13} Kromer, M., Fink, M., 
         Stanishev, V., et al. 2013, \mnras, 429, 2287
\bibitem[Larsson et al.(2013)]{Larsson13} Larsson, J., Fransson, C., Kjaer, K., 
         et al.\ 2013, \apj, 768, 89 
\bibitem[Li et al.(2003)]{li03} Li, W., Filippenko, A. V., Chornock, R., 
         et al. 2003, \pasp, 115, 453L
\bibitem[Li et al.(2011)]{li11} Li W., Leaman, J., Chornock, R., et al., 2011, \mnras, 412, 1441
\bibitem[Livne(1999)]{Livne99} Livne, E.\ 1999, \apjl, 527, L97
\bibitem[Livne et al.(2005)]{Livne05} Livne, E., Asida, S. M., H{\"o}flich, P.\ 2005, \apj, 632, 443
\bibitem[Maeda et al.(2010)]{Maeda2010} Maeda, K., Taubenberger, S., 
         Sollerman, J., et al.\ 2010, \apj, 708, 1703 
\bibitem[McConnachie et al.(2005)]{McConn2005} McConnachie, A.~W., 
         Irwin, M.~J., Ferguson, A.~M.~N., et al.\ 2005, \mnras, 356, 979 
\bibitem[Morton(1991)]{Morton91} Morton, D.~C.\ 1991, \apjs, 77, 119 
\bibitem[Niemeyer \& Hillebrandt(1995)]{Neimeyer95} Niemeyer, J.~C., \& 
         Hillebrandt, W.\ 1995, \apj, 452, 779 
\bibitem[Niemeyer et al.(1996)]{Niemeyer1996} Niemeyer, J.~C., 
         Hillebrandt, W., \& Woosley, S.~E.\ 1996, \apj, 471, 903 
\bibitem[Nomoto(1982)]{Nomoto82} Nomoto, K.\ 1982, \apj, 253, 798 
\bibitem[Nomoto et al.(1984)]{Nomoto84} Nomoto, K., Thielemann, F.-K., \& 
         Yokoi, K.\ 1984, \apj, 286, 644 
\bibitem[Nomoto et al.(2003)]{nomoto03} Nomoto, K., Uenishi, T., 
        {Kobayashi}, C. et al.\ 2003, in: {\em From Twilight to Highlight: 
         The Physics of Supernovae}, p. 115
\bibitem[Nugent et al.(2011)]{Nugent2011} Nugent, P.~E., Sullivan, M., 
         Cenko, S.~B., et al.\ 2011, \nat, 480, 344 
\bibitem[Osterbrock(2001)]{Osterbrock2001} Osterbrock, D.~E.\ 2001, 
         `Walter Baade A Life in Astrophysics',  Donald E.~Osterbrock, 
         Princeton, NJ, USA: Princeton University Press. 2001, p. 93 
\bibitem[Pakmor et al.(2011)]{Pakmor11} Pakmor, R., Hachinger, S., 
         R{\"o}pke, F.~K., \& Hillebrandt, W.\ 2011, \aap, 528, A117 
\bibitem[Pastorello et al.(2008)]{Pastorello08} Pastorello, A., et al.\ 2008, \mnras, 389, 113
\bibitem[Penney et al.(2014)]{Penney2014} Penney, R., H{\"o}flich, P., 2014, \apj 795, 84 
\bibitem[Remming et al.(2014)]{Remming2014} Reming, I.S., Khokhlov, A.M. 2014, \apj 794, 87
\bibitem[Perets et al.(2011)]{Perets2011} Perets, H.~B., Badenes, 
         C., Arcavi, I., Simon, J.~D., \& Gal-yam, A.\ 2011, \apj, 730, 89 
\bibitem[Plewa(2007)]{Plewa07} Plewa, T.\ 2007, \apj, 657, 942 
\bibitem[Quimby et al.(2006)]{Quimby06} Quimby, R., H{\"o}flich, P., Kannappan, S.~J., 
         et al.\  2006, \apj, 636, 400 
\bibitem[Rasio \& Shapiro(1994)]{rasio94} Rasio, F.~A., \& Shapiro, S.~L.\ 1994, \apj, 432, 242 
\bibitem[Reinecke et al.(1999)]{Reinecke99} Reinecke, M., Hillebrandt, W., \& 
         Niemeyer, J.~C.\ 1999, \aap, 347, 739 
\bibitem[Reinecke et al.(2002)]{r02} Reinecke, M., Hillebrandt, W., \& Niemeyer, J.~C.\ 2002, \aap, 391, 1167 
\bibitem[R{\"o}pke et al.(2006)]{roepke06} R{\"o}pke, F.~K., Hillebrandt, W., 
         Niemeyer, J.~C., \& Woosley, S.~E.\ 2006, \aap, 448, 1 
\bibitem[R{\"o}pke et al.(2007)]{roepke07} R{\"o}pke, F.K., Woosley, S.E, 
         Hillebrandt, W. \ 2007, \apj, 660, 1344
\bibitem[R{\"o}pke et al.(2012)]{roepke12} {R{\"o}pke}, F.K., Kromer, D., 
         {Seitenzahl}, I.R. et al. 2012, ApJLet 750, L19
\bibitem[Seitenzahl et al.(2013)]{seitenzahl13} Seitenzahl, I.~R., Ciaraldi-Schoolmann, F., 
         R{\"o}pke, F.~K., et al.\ 2013, \mnras, 429, 1156 
\bibitem[Sjouwerman \& Dickel(2001)]{SD01} Sjouwerman, L.~O., \& Dickel, J.~R.\ 2001, 
         Young Supernova Remnants, 565, 433 
\bibitem[Di Stefano et al.(2011)]{Stephano11} Di Stefano, R., 
         Voss, R., \& Claeys, J.~S.~W.\ 2011, \apjl, 738, LL1 
\bibitem[Di Stefano \& Kilic(2012)]{Stefano2012} Di Stefano, R., \& Kilic, M.\ 2012, \apj, 759, 56 
\bibitem[Stritzinger et al.(2014)]{Stritzinger2014} Stritzinger, M.~D., Hsiao, E., 
        Valenti, S., et al.\ 2014, \aap, 561, AA146 
\bibitem[Sugimoto \& Nomoto(1980)]{Sugimoto80} Sugimoto, D., \& Nomoto, K.\ 1980, \ssr, 25, 155 
\bibitem[Timmes \& Woosley(1992)]{Timmes1992} Timmes, F.~X., \& 
         Woosley, S.~E.\ 1992, \apj, 396, 649 
\bibitem[Wang \& Han(2012)]{wang2012} Wang, B., \& Han, Z.\ 2012, New Astronomy, 56, 122 
\bibitem[Webbink(1984)]{webbink84} Webbink, R.~F.\ 1984, \apj, 277, 355 
\bibitem[Yamaoka et al.(1992)]{yamaoka92} Yamaoka, H., Nomoto, K., 
         Shigeyama, T., \& Thielemann, F.-K.\ 1992, \apjl, 393, L55 
\end{thebibliography}
\end{document}